\documentclass[twocolumn,pra,letterpaper,showpacs,citesort]{revtex4}
\usepackage[caption=false]{subfig}
\usepackage{geometry,amsthm,amsmath,amssymb,hyperref,graphicx}
\input{Qcircuit}
\def\be{\begin{eqnarray}}
\def\ee{\end{eqnarray}}

\renewcommand{\ket}[1]{{| #1 \rangle}}
\newcommand{\eps}{\varepsilon}
\newcommand{\epsCSS}{\varepsilon_{\text{\sc css}}}
\newcommand{\epsin}{\varepsilon_{\text{in}}}

\newcommand{\MeasZ}{\mathcal{M}_{Z}}

\newcommand{\MeasZZ}{\mathcal{M}_{ZZ}}
\newcommand{\MeasZZZ}{\mathcal{M}_{ZZZ}}
\newcommand{\MeasX}{\mathcal{M}_{X}}

\newcommand{\Prep}{\mathcal{P}}
\newcommand{\Meas}{\mathcal{M}}
\newcommand{\PrepPlus}{\Prep_{\ket{+}}}

\newcommand{\PrepZero}{\Prep_{\ket{0}}}

\newcommand{\PrepCat}{\Prep_{\ket{\mbox{cat}}}}
\newcommand{\PrepZZCat}{\Prep_{\ket{ZZ\mbox{-cat}}}}
\newcommand{\PrepZZZCat}{\Prep_{\ket{ZZZ\mbox{-cat}}}}
\newcommand{\Perr}{P_{\text{err}}}
\newcommand{\Gfund}{\mathcal{G}_{\text{fund}}}
\newcommand{\GCSS}{\mathcal{G}_{\text{CSS}}}

\newcommand{\Inject}{\mathcal{J}}

\renewcommand{\control}{*!<0em,.025em>-=-<.4em>{\bullet}} 

\begin{document}
\title{Fault-tolerant quantum computation \\with asymmetric Bacon-Shor codes}
\author{Peter Brooks and John Preskill}
\affiliation{Institute for Quantum Information, California Institute of Technology, Pasadena, CA 91125, USA}

\begin{abstract}
We develop a scheme for fault-tolerant quantum computation based on asymmetric Bacon-Shor codes, which works effectively against highly biased noise dominated by dephasing. We find the optimal Bacon-Shor block size as a function of the noise strength and the noise bias, and estimate the logical error rate and overhead cost achieved by this optimal code. Our fault-tolerant gadgets, based on gate teleportation, are well suited for hardware platforms with geometrically local gates in two dimensions. 
\end{abstract}

\maketitle

\section{Introduction}

The theory of fault-tolerant quantum computation \cite{shor1996fault,gottesman2009introduction} has established that noisy quantum computers can operate reliably provided the noise is neither too strong nor too strongly correlated. In a fault-tolerant quantum circuit, carefully designed gadgets process logical qubits protected by quantum error-correcting codes. 

Typical fault-tolerant gadgets are designed to work effectively against generic noise without any special structure. But in some physical settings, the noise is expected to be highly biased, with dephasing in the computational basis far more likely than bit flips. This paper addresses how noise bias can be exploited to improve the reliability of fault-tolerant quantum circuits.

Specifically, we have in mind a setting in which, to an excellent approximation, the computational basis states $\{|0\rangle,|1\rangle\}$ are the energy eigenstates for an unperturbed qubit, and single-qubit gates which are diagonal in this basis can be performed by adiabatically adjusting the energy splittings between these states. Similarly, diagonal two-qubit gates are performed by adjusting the energy splittings of the four (approximate) energy eigenstates $\{|00\rangle,|01\rangle,|10\rangle,|11\rangle\}$. Various noise sources may induce fluctuations in these energy spacings, and pulse imperfections may cause gates to be over-rotated or under-rotated; in either case the noisy gate deviates from the ideal gate, but remains diagonal. Other physical processes may induce transitions between energy eigenstates, but we will assume that this non-diagonal noise is far weaker than the dominant diagonal noise.

Our scheme for fighting biased noise is based on asymmetric Bacon-Shor codes \cite{Shor95,Bacon06-2,
napp2012optimal}. These quantum codes combine together a length-$m$ repetition code protecting against dephasing and a length-$n$ repetition code protecting against bit flips; by ``asymmetric'' we mean $m > n$, so the code works more effectively against dephasing than against bit flips. A complete universal set of fault-tolerant gates can be constructed using only diagonal gates, plus single-qubit measurements in the $X$ eigenstate basis $\{|\pm\rangle =\frac{1}{\sqrt{2}}\left(|0\rangle \pm |1\rangle \right)\}$ and preparation of single-qubit $|+\rangle$ states. 

In most previous studies of quantum fault-tolerance based on Bacon-Shor codes, these codes have been concatenated to build a hierarchy of codes within codes \cite{aliferis2007subsystem,Cross07}. Here we will mainly study the performance of a single large block code rather than a concatenated scheme. For the Bacon-Shor code family, used without concatenation, there is no accuracy threshold; rather, for a fixed value of the noise strength and noise bias, there is an optimal block size which achieves the best performance.

Our gadget constructions and analysis extend the results in \cite{Aliferis08,Aliferis09}, where the case $n=1$ was regarded as the bottom layer of a concatenated code. In contrast to \cite{Aliferis08,Aliferis09}, we consider the case where the gates are required to be geometrically local in a two-dimensional array. In the geometrically local case, logical blocks can be transported as needed via teleportation, with the fundamental operations still limited to diagonal gates, $X$ measurements, and $|+\rangle$ preparations. 

We have obtained upper bounds on the optimal performance and overhead cost of logical gates by deriving analytic formulas, and our estimates are far from tight. We expect that significantly better results could be obtained using numerical simulations assuming independent biased noise, which we have not attempted. In particular, our analysis includes quite conservative estimates of the probability of failure for preparations of cat states that are used for measurements of logical Pauli operators.

Another particularly attractive approach to achieving quantum fault tolerance with geometrically local gates is based on topological codes \cite{dennis2001topological,raussendorf2007topological,fowler2009high}, which unlike Bacon-Shor codes have an accuracy threshold and hence can reach arbitrarily low error rates per logical gate. Topological codes can be adapted for optimized protection against biased noise, and they perform well against biased noise even without any such adaptation. Our current results do not conclusively identify noise parameter regimes for which Bacon-Shor codes are clearly superior to topological codes. However, our scheme has several appealing features --- for example, the optimal error rate per logical gate is achieved with a relatively modest number of physical qubits per code block, and is not too adversely affected when qubit measurements are noisier than quantum gates. Furthermore, only relatively modest classical computational resources are needed to interpret error syndromes. 

We review Bacon-Shor codes in Sec. II, describe our biased noise model in Sec. III, and construct our fault-tolerant gadgets in Sec. IV. We obtain upper bounds on the performance of our fault-tolerant logical CNOT gate in Sec. V, repeat the analysis in Sec. VI for the case where two-qubit gates are required to be geometrically local, and report numerical values in Sec. VII. In Sec. VIII and IX we discuss the state injection and state distillation procedures needed to complete our fault-tolerant  universal gate set. 

Fault-tolerant gadgets protecting against biased noise have also been discussed previously in \cite{Gourlay00,Evans07,Stephens08}. 

\section{Bacon-Shor codes}

Bacon-Shor codes are quantum subsystem codes which are constructed by combining together two quantum repetition codes, one protecting against $Z$ (phase) errors and the other protecting against $X$ (bit flip) errors. We will consider using Bacon-Shor codes to protect quantum information against biased noise, such that $Z$ errors are much more common than $X$ errors; therefore, the length $m$ of the code protecting against $Z$ errors will be longer than the length $n$ of the code protecting against $X$ errors. In this case we say that the code is asymmetric. The code protects one encoded qubit in a block of $nm$ physical qubits.

It is convenient to arrange the physical qubits in a rectangular lattice with $n$ rows and $m$ columns. The code has $(m-1)$ $Z$-type stabilizer generators, or check operators, which can be measured to determined a syndrome for detecting $X$ errors, and $n-1$ $X$-type stabilizer generators, which can be measured to determine the syndrome for $Z$ errors. Each $Z$-type check operator is a weight-$2m$ operator applying $Z$ to each qubit in a pair of adjacent rows, and each $X$-type check operator is a weight-$2n$ operator applying $X$ to each qubit in a pair of adjacent columns. Because each $Z$-type check operator ``collides'' with each $X$-type check operator at four lattice sites, the $Z$-type and $X$-type check operators are mutually commuting and can be measured simultaneously. The logical Pauli operator $Z^L$ acting on the encoded qubit is a tensor product $Z^{\otimes m}$ acting on all the qubits in a (long) row; all rows are equivalent because a product of two such row operators is in the code stabilizer. The operator $Z^L$ collides with each $X$-type check operator at two sites; therefore it commutes with these check operators and hence preserves the code space. The logical Pauli operator $X^L$ is a tensor product $X^{\otimes n}$ acting on all qubits in a (short) column; again the columns are equivalent, and $X^L$ commutes with the $Z$-type check operators. The operators $Z^L$ and $X^L$ collide at a single site and hence anticommute.

\begin{figure}[h]
 	\begin{center}
 	\includegraphics[height=1.25in]{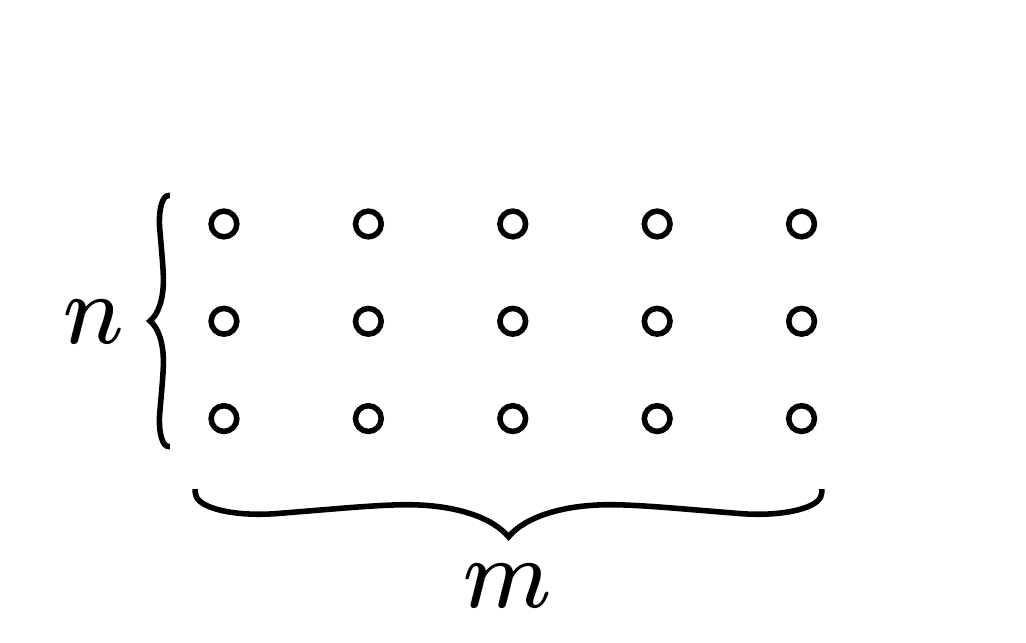}
 	\end{center}
	\caption{Bacon-Shor qubits arranged in an $n \times m$ lattice.}
	\label{fig:code-block}
\end{figure}

Assuming that $m$ and $n$ are both odd, the code can correct $(m-1)/2$ $Z$ errors and $(n-1)/2$ $X$ errors. To perform a decoded $X^L$ measurement, we can measure all $nm$ qubits in the $X$ basis, compute the parity of the outcomes for each of the $m$ columns, and then perform a majority vote on the $m$ column parities. The decoded measurement fails only if $(m+1)/2$ columns each have at least one $Z$ errors. Similarly, we can perform a decoded $Z^L$ measurement by measuring all qubits in the $Z$ basis, computing the parity of the outcomes for each of the $n$ rows, and then performing a majority vote on the $n$ row parities. This decoded measurement fails only if $(n+1)/2$ rows each have at least one $X$ error.

The code also has a gauge algebra of Pauli operators that commute with the check operators and with the logical operators, though the gauge operators do not necessarily commute with one another. The $Z$-type gauge algebra is generated by weight-two operators with $Z$ acting on a pair of neighboring qubits in the same column, and the $X$-type gauge algebra is generated by weight-two operators with $X$ acting on a pair of neighboring qubits in the same row. Thus each $X$-type check operator is a product of $n$ weight-two gauge operators, and each $Z$-type check operator is a product of $m$ weight-two gauge operators. Conveniently, then, the error syndrome can be determined by measuring only weight-two gauge operators, and furthermore each of these gauge operators is geometrically local, acting on a pair of neighboring qubits. Measuring an $X$-type gauge operator disturbs the values of some $Z$-type gauge operators (and vice versa), but without inflicting any damage on the protected logical subsystem.

\begin{figure}[h]
 	\begin{center}
 	\subfloat{\includegraphics[width=0.4\textwidth]{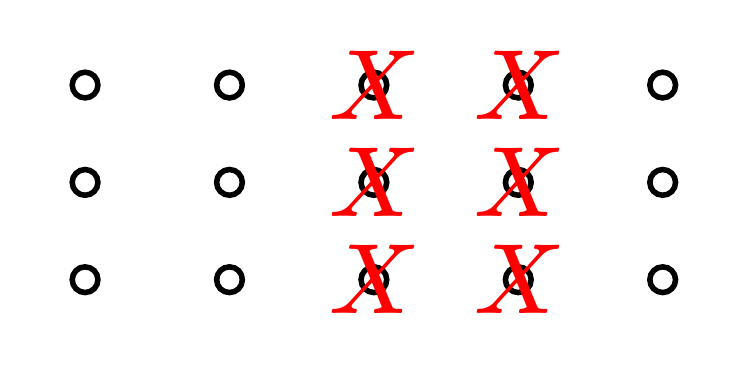}} \\
	\subfloat{\includegraphics[width=0.4\textwidth]{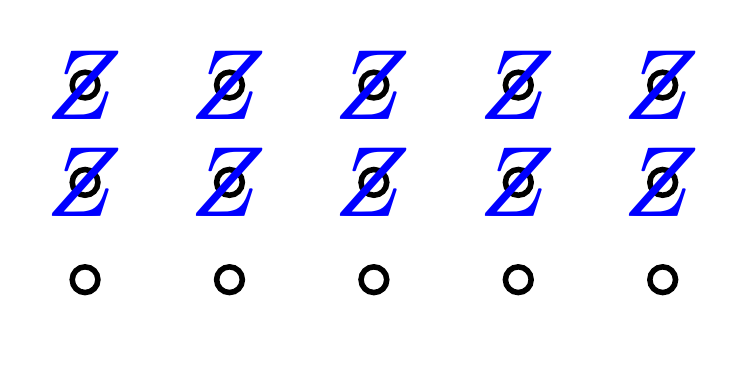}}
 	\end{center}
	\caption{(Color online) $X$-type stabilizer operator (top) and $Z$-type stabilizer operator (bottom)}
	\label{fig:stabilizers}
\end{figure}

\section{Noise model}

We consider a local stochastic biased noise model, in which we distinguish between a rate $\eps$ for dephasing faults and a (perhaps much lower) rate $\eps'$ for general faults in diagonal gates. The ratio $R = \eps / \eps'$ will be called the noise bias. For simplicity we will assume that $\eps$ is also the error rate for single-qubit preparations and measurements in most of the formulas we display, though we will treat the case where preparations and measurements are noisier than diagonal gates in Sec.~\ref{subsec:measurement-bias}. A dephasing fault, which we will also call a diagonal fault, is a trace-preserving completely positive map such that all Kraus operators are diagonal in the computational basis; for a general fault, which we will also call a non-diagonal fault, the Kraus operators are unrestricted. We assume that the sum of the probabilities for all fault paths with $r$ dephasing faults and $s$ non-dephasing faults at specified circuit locations is no greater than $\eps^r (\eps')^s$. 

\section{Fault-tolerant gadgets}

The construction of fault-tolerant gadgets for universal quantum computation follows Ref. \cite{Aliferis08}. We implement the logical gate set
\begin{equation}
	\GCSS^L = \{ \PrepPlus^L, \PrepZero^L, \MeasX^L, \MeasZ^L, \mbox{CNOT}^L \}
\end{equation}
using the physical gate set
\begin{equation}\label{eq:Gfund}
	\Gfund = \{ \PrepPlus, \MeasX, \mbox{CZ} \};
\end{equation}
here $\MeasX$, $\MeasZ$ denote measurements of the Pauli operators $X$ and $Z$, $\PrepPlus$, $\PrepZero$ denote preparations of the $X$ and $Z$ eigenstates with eigenvalue $1$, and CZ denotes the two-qubit controlled phase gate
\begin{equation}
\mbox{CZ} = \exp\left( -i \frac{\pi}{4}\left(Z-I\right)\otimes\left(Z-I\right)\right),
\end{equation}
which is diagonal in the computational ($Z$-eigenstate) basis $|00\rangle,|01\rangle,|10\rangle,|11\rangle$ with eigenvalues $(1,1,1,-1)$. A diagonal gate like this one can be realized by adiabatically perturbing the energy splittings of the computational basis states, and hence could plausibly have highly biased noise. Note that we have not included $\MeasZ$ in the physical gate set Eq.~(\ref{eq:Gfund}), because it will not be needed in our fault-tolerant gadget constructions. 

The CZ gate acts symmetrically on the two qubits, and, like any Clifford gate, it can be usefully characterized by how it propagates Pauli errors:
\begin{eqnarray}\label{eq:cz-rule}
{\rm CZ}: XI & \rightarrow XZ, \quad IX & \rightarrow ZX, \nonumber\\
ZI & \rightarrow ZI, \quad IZ & \rightarrow IZ.
\end{eqnarray}
Of course, CZ commutes with $Z$ acting on either qubit, and propagates an $X$ acting on either qubit to a $Z$ error acting on the other. These properties are nice if $Z$ errors dominate over $X$ errors and we use a code that corrects $Z$ errors more effectively than $X$ errors. The $CZ$ gates do not propagate the relatively common $Z$ errors, and though $X$ errors do propagate they are transformed to $Z$ errors, which the code handles more readily.

To achieve universal quantum computation, we supplement the logical gate set $\GCSS^L$ with high-fidelity encoded states $\ket{i}= \frac{1}{\sqrt{2}}\left(|0\rangle + i|1\rangle\right)$ and $\ket{T}= \frac{1}{\sqrt{2}}\left(|0\rangle + e^{i\pi/4}|1\rangle\right)$, which can be prepared using state injection and distilled using the $\GCSS^L$ operations \cite{Bravyi08}. Gates that complete a universal set can then be constructed using these states and the $\GCSS^L$ gates \cite{Knill05}.


\subsection{\texorpdfstring{$\MeasX^L$}{Measure X} gadget} 
To perform a destructive $\MeasX^L$ measurement we first perform independent $\MeasX$ measurements on all qubits in the code block, and then compute the parity of the $n$ measurement outcomes in each column. Finally a majority vote of the $m$ column parities yields the logical measurement outcome. The result agrees with an ideal measurement if the sum of the number of $Z$ errors in the block and the number of faulty $\MeasX$ measurements is no larger than $(m-1)/2$.

\subsection{\texorpdfstring{$\MeasZ^L$}{Measure Z} gadget} 
If $\MeasZ$ were included in our repertoire of physical gates, then we could perform a destructive $\MeasZ^L$ using the dual of the procedure described above for the destructive $\MeasX^L$. That is, we could perform independent $\MeasZ$ measurements on all qubits in the code block, compute the parity of the $m$ measurement outcomes in each row, and perform a majority vote of the $n$ row parities to obtain the logical measurement outcome. The result agrees with an ideal measurement if the sum of the number of $X$ errors in the block and the number of faulty $\MeasZ$ measurements is no larger than $(n-1)/2$.

However, for our fault-tolerant gadget constructions we will require instead a {\em non-destructive} $\MeasZ^L$ measurement, such that the ideal measurement procedure leaves encoded $Z^L$ eigenstates intact. In principle a non-destructive measurement of the row parity could be done by preparing a single ancilla qubit in the state $|+\rangle$, performing $m$ CZ gates in succession acting on the ancilla qubit and the $m$ qubits in the row, and finally measuring the ancilla qubit in the $X$ basis. However, this procedure is not fault-tolerant for two reasons. First, a single $Z$ error acting on the ancilla qubit can flip the measurement outcome. Second, a single $X$ error acting on the ancilla can propagate multiple times, producing a high-weight $Z$ error acting on the qubits in the row. 

This second problem can be addressed by replacing the single-qubit ancilla by an $m$-qubit cat state $\ket{+^\text{cat}} = \frac{1}{\sqrt{2}} \left(\ket{0}^{\otimes m} + \ket{1}^{\otimes m}\right)$, so that each ancilla qubit interacts via a CZ gate with only one data qubit, limiting the error propagation. The fault-tolerant preparation of the cat state is discussed below. After the $m$ CZ gates, the cat state is read out in the basis  $\ket{\pm^\text{cat}} = \frac{1}{\sqrt{2}} \left(\ket{0}^{\otimes m} \pm \ket{1}^{\otimes m}\right)$ to determine the row parity. This measurement of $X^{\otimes m}$ is performed destructively by measuring each of the $m$ ancilla qubits in the $X$ basis and computing the parity of the outcomes.

To address the first problem (that a single fault can flip the measured parity), the parity measurement is repeated $r$ times for each row, each time with a fresh cat state, and the majority of the results is computed. The repetition provides protection against $Z$ errors in the ancilla, but of course there may also be (relatively rare) $X$ errors acting on the data qubits due to faults during the $Z^L$ measurement circuit. These faults cause a logical $X^L$ error only if $X$ errors occur in a majority of the $n$ rows. 

Actually, we find that, depending on the rates for diagonal and non-diagonal faults, it may be advantageous to use a shorter ancilla prepared in a cat state of length $p$, where $1 \leq p < m$. In that case some ancilla qubits interact with more than one data qubit, increasing the danger of error propagation, but on the other hand errors are less likely to occur during the preparation of the (shorter) cat state. However, using a length-$m$ cat state has the significant advantage that the $\MeasZ^L$ measurement can be conveniently executed using geometrically local gates on the two-dimensional lattice, as we will discuss in Sec.~\ref{sec:local}.

This procedure for measuring $Z^L$ can easily be extended to build gadgets that perform the parity measurements $\MeasZZ^L$, $\MeasZZZ^L$ on multiple logical qubits. 

\begin{figure*}[ht]
 	\begin{center}
 	\includegraphics[height=4in]{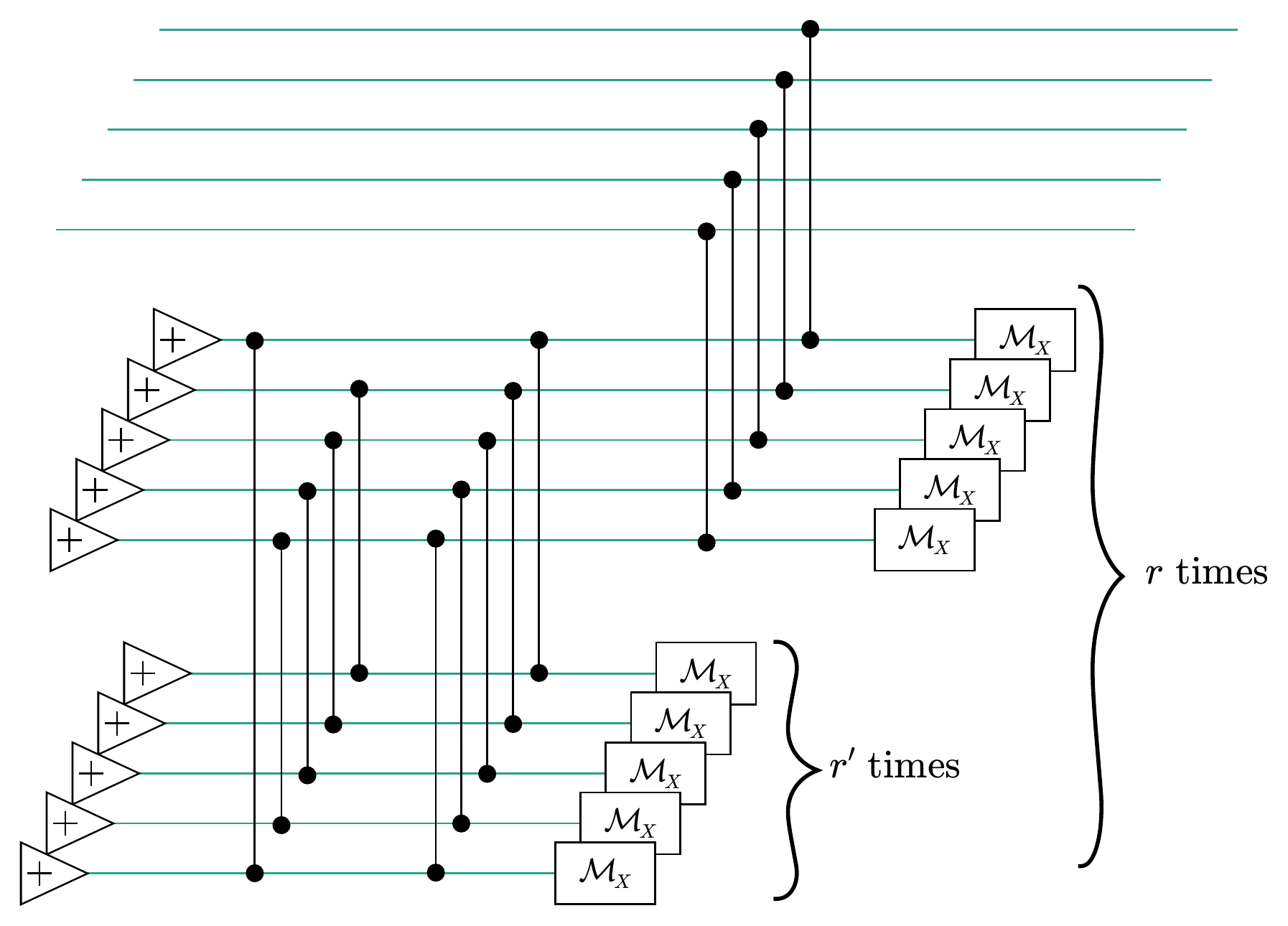}
 	\end{center}
	\caption{(Color online) Gadget for a single row of the non-destructive $\MeasZ^L$ measurement, where the cat state ancilla length $p$ is equal to the row length $m$.}
	\label{fig:meas-Z-row}
\end{figure*}

\subsection{Cat state preparation}
The length-$p$ cat state $\ket{+^\text{cat}} = \frac{1}{\sqrt{2}} \left(\ket{0}^{\otimes p} + \ket{1}^{\otimes p}\right)$ may be characterized as the simultaneous eigenstate with eigenvalue 1 of $X^{\otimes p}$ and of $p-1$ $ZZ$ operators, each one acting on a pair of neighboring qubits. To prepare the cat state, we start by preparing the product state $|+\rangle^{\otimes p}$ (an eigenstate of $X^{\otimes p}$), and proceed to measure the $p-1$ $ZZ$ operators (each of these measurements is executed using yet another ancilla qubit). If every measurement has the outcome $ZZ=+1$, then the cat state has been prepared successfully, assuming the measurements are flawless. Otherwise, the measurement outcomes provide a syndrome pointing to $X$ errors in the cat state (each $ZZ=-1$ outcome identifies the endpoint of a chain of $X$ errors). These $X$ errors in the cat state need not be corrected; instead we can keep track of their propagation as the computation proceeds. Specifically, the $X$ errors in the cat state propagate to become $Z$ errors acting on the measured row of the data block, which should be taken into account when we decode a later $X^L$ measurement of that block. Following Knill  \cite{Knill05}, we say that the measured syndrome is used to update the ``Pauli frame'' of the computation. 

To improve robustness, we add one more (redundant) $ZZ$ measurement acting on the first and last qubit in the cat state. Then an even number among the $p$ $ZZ$ measurements should have the outcome $-1$ if all measurements are correct; if in fact we find an odd number of $-1$ outcomes than we reject the syndrome. (Adding further redundant checks would further improve the reliability of the syndrome extraction, though we will not include these.) We conduct $r'$ rounds of syndrome measurement, including rejected rounds, so that the number of accepted rounds may be fewer than $r'$. 

In principle, the cat state Pauli frame could be determined by performing a perfect matching on a two-dimensional graph representing the space-time history of the cat state syndrome measurement \cite{dennis2001topological}. Instead, we will consider a much simpler and less effective method for decoding the syndrome history. Though far from optimal, our procedure has two advantages over the perfect matching algorithm --- it requires only very modest (classical) computational resources compared to perfect matching, and it is also much easier to study analytically. 

In our unsophisticated scheme, we determine the Pauli frame using whichever syndrome occurs most frequently among the accepted rounds, which we call the ``winning'' syndrome. (Our analysis does not depend on what rule is used when two or more syndromes tie for the distinction of being most frequent, since we pessimistically assume that the preparation gadget fails catastrophically in the event of such a tie.) The state preparation protocol is deterministic in the sense that ancillas are never discarded, and the number of rounds of syndrome measurement is always $r'$ irrespective of the measurement outcomes.

The complete circuit for measuring the row parity, including repeated measurement of the cat state syndrome, is shown in \autoref{fig:meas-Z-row}. In the circuit shown, the cat state is prepared first, and then used to measure the parity of a row. Since the CZ gates commute with each other, we could alternatively perform the $ZZ$ measurements on the cat state \emph{after} the ancilla interacts with the data. 

Even if the cat state is ``prepared'' after its ``use,'' $X$ errors in the ancilla during the preparation can still cause trouble. The $X$ errors cause the cat state syndrome to evolve as it is measured repeatedly, and therefore the syndrome that occurs most often, even if valid when measured, may differ from the syndrome that is applicable when the cat state interacts with the data. The effect of this syndrome ``drift'' needs to be included in our error analysis.

\subsection{\texorpdfstring{$\PrepPlus^L$}{Plus preparation} gadget} The logical state $\ket{+}^L$, for a particular ``choice of gauge,'' is a product of $m$ length-$n$ cat states, one for each column of the block. This state is an eigenstate with eigenvalue $1$ of $X^L=X^{\otimes n}$ acting on any column, and of the gauge operator $ZZ$ acting on a pair of neighboring qubits in any column (and hence also a +1 eigenstate of all the $Z$-type check operators). We prepare these $m$ length-$n$ cat states in parallel, using the same procedure for preparing length-$p$ cat states described above. In this case we repeat the syndrome measurement for each column $r_+$ times. 


\subsection{Error correction gadget} 
\label{subsec:ec}

\begin{figure}[ht]
 	\begin{center}
 	\includegraphics[width=0.45\textwidth]{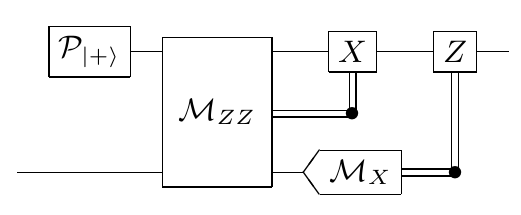}
 	\end{center}
	\caption{One-bit teleportation circuit.}
	\label{fig:teleportation}
\end{figure}

Error correction can be performed on a logical block using the ``one-bit teleportation'' gadget shown in \autoref{fig:teleportation}. This circuit transfers the single-qubit state $|\psi\rangle$ from qubit 2 to qubit 1. We prepare qubit 1 in the state $|+\rangle =\frac{1}{\sqrt{2}}(\left |0\rangle + |1\rangle\right)$, perform the joint measurement $\mathcal{M}_{Z^1Z^2}$ on the two qubits, then perform the single-qubit measurement $\mathcal{M}_{X^2}$. After applying Pauli operators conditioned on the two measurement outcomes, the result is that the input state $|\psi\rangle$ of qubit 2 is moved to qubit 1. The conditional $X$ and $Z$ gates in this teleportation gadget need not actually be executed; instead the Pauli frame is updated to reflect the measurement outcomes. One-bit teleportation can be performed at the logical level using the fault-tolerant measurement gadgets $\MeasZZ^L$ and $\MeasX^L$ that we have already described.

\subsection{\texorpdfstring{CNOT$^L$}{CNOT} gadget} 
\label{subsec:cnot}
\begin{figure*}[t]
 	\begin{center}
 	\includegraphics[width=0.9\textwidth]{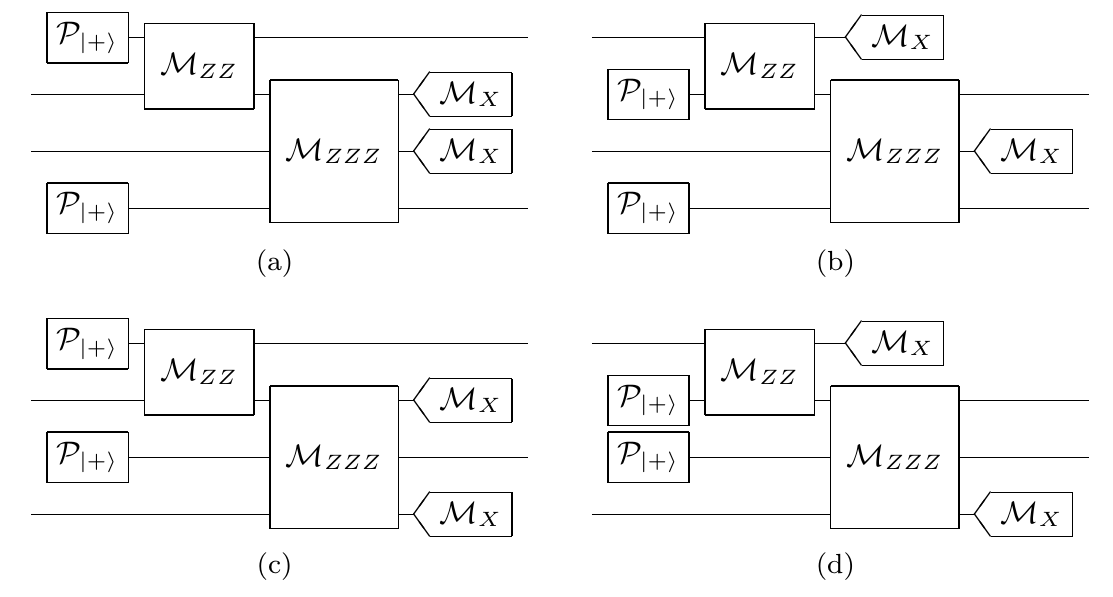}
 	\end{center}
	\caption{Four versions of the CNOT gadget.}
	\label{fig:CNOT-gadget}
\end{figure*}

The gadget shown in \autoref{fig:CNOT-gadget} applies a fault-tolerant ``teleported'' CNOT$^L$ logical operation, up to a known Pauli error determined by the measurement outcomes \cite{Aliferis08}. Aside from executing the logical gate, the gadget also achieves error correction by measuring both incoming blocks and preparing fresh outgoing blocks.

\begin{figure}[h]
 	\begin{center}
 	\includegraphics[width=0.45\textwidth]{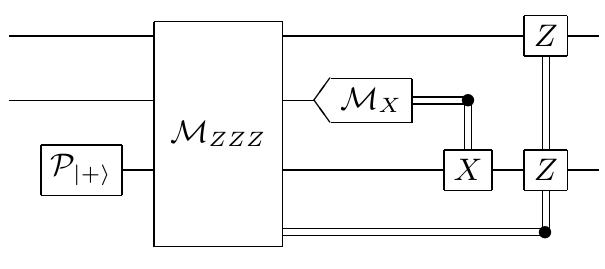}
 	\end{center}
	\caption{Teleported CNOT gate.}
	\label{fig:teleported-CNOT}
\end{figure}

To better understand how the gadget works, it is helpful to divide it into two parts. One part is the ``one-bit teleportation'' gadget described in Sec.~\ref{subsec:ec}, which performs error correction on a single block. The other component of the CNOT gadget is shown in \autoref{fig:teleported-CNOT}, where again the Pauli frame updates are explicitly indicated. We may verify that this circuit realizes a CNOT gate by checking that it propagates Pauli operators according to the rule
\begin{eqnarray}\label{eq:cnot-rule}
{\rm CNOT}: XI & \rightarrow XX, \quad IX & \rightarrow IX, \nonumber\\
ZI & \rightarrow ZI, \quad IZ & \rightarrow ZZ.
\end{eqnarray}
We denote the measurement outcomes of $\Meas_X$ and $\Meas_{ZZZ}$ as $(-1)^a$ and $(-1)^b$ respectively. Since $|+\rangle$ is an eigenstate of $X$ with eigenvalue $+1$, $XI$ acting on the input qubits is equivalent to $XIX$ acting on the three qubits in the circuit, which commutes with $\Meas_{ZZZ}$ and hence becomes $XX$ acting on the output qubits. Likewise, $IX$ acting on the input qubits is equivalent to $IXX$, which commutes with $\Meas_{ZZZ}$ and becomes $(-1)^a IX$ acting on the output qubits using the measurement outcome $IXI=(-1)^a$. $ZI$ acting on the input qubits commutes with $\Meas_{ZZZ}$ and trivially maps to $ZI$ acting on the output qubits, while $IZ$ acting on the input qubits becomes $IZI$, which maps to $(-1)^b ZIZ$ using the measurement outcome $ZZZ=(-1)^b$. Hence we have found
\begin{eqnarray}
{\rm CNOT}: XI & \rightarrow XX, \quad IX & \rightarrow (-1)^a IX, \nonumber\\
ZI & \rightarrow ZI, \quad IZ & \rightarrow (-1)^b ZZ ,\nonumber\\
\end{eqnarray}
which is corrected to Eq.(\ref{eq:cnot-rule}) by applying $Z^a$ to both output qubits and $X^b$ to the second qubit. When we combine the CNOT gate with one-bit teleportation of the first qubit, the measurements $\Meas_{ZZ}$ and $\Meas_{ZZZ}$ commute and hence can be done in either order.

\newcommand{\blah}[1]{\control \cwx[#1] \cw}

Four different versions of the CNOT gadget are shown in \autoref{fig:CNOT-gadget} Let us denote the blocks appearing in the figure as blocks 1, 2, 3, 4, counting from the top down. Version $(a)$ of the CNOT gadget contains the preparations and measurements $\PrepPlus[1]$, $\PrepPlus[4]$, $\MeasZZ[12]$, $\MeasZZZ[234]$, $\MeasX[2]$, $\MeasX[3]$. This gadget has some arbitrary features which are exploited to construct the other versions. 
Since $ZZ[12]ZZZ[134] = ZZZ[234]$, we extract the same parity information if $\MeasZZZ[234]$ is replaced by $\MeasZZZ[134]$; also, since $\MeasZZ$ acts symmetrically on its two measured blocks, we may permute the blocks, in effect replacing $\PrepPlus[1]$ and $\MeasX[2]$ by $\PrepPlus[2]$ and $\MeasX[1]$, thus obtaining version $(b)$. Similarly, since $\MeasZZZ$ acts symmetrically, we may replace $\PrepPlus[4]$ and $\MeasX[3]$ by $\PrepPlus[3]$ and $\MeasX[4]$, thus obtaining versions $(c)$ and $(d)$. In all these realizations of the gadget, the upper incoming block is the control block and the lower incoming block is the target block; by replacing $\MeasZZ[12]$ and $\MeasZZZ[234]$ by $\MeasZZZ[123]$ and $\MeasZZ[34]$ we can reverse the order of the CNOT. 

This freedom to choose the incoming control block to be either block 1 or block 2, and to choose the incoming target block to be either block 3 or block 4, will be useful when we consider in Sec.~\ref{sec:local} how to realize the CNOT gadget using geometrically local gates. No matter how we configure the gadget, one block participates in both $\MeasZZ$ and $\MeasZZZ$. However, for versions $(a)$ and $(c)$ in \autoref{fig:CNOT-gadget}, the twice measured block is an incoming (control) data block, while for versions $(b)$ and $(d)$ the twice measured block is an outgoing (control) data block. 

The CNOT$^L$ gadget is characterized by four parameters: the length $p$ of the cat states used in the $Z$-type measurements  and the number of times measurements are repeated ($r_+$ for the $|+\rangle^L$ preparation, $r'$ for the cat state preparation, and  $r$ for $ZZ^L$ and $ZZZ^L$ measurements). In principle we could choose different values of these parameters in the measurements $\MeasZZ^L$ and $\MeasZZZ^L$, but in practice we have found it optimal or near optimal to use the same parameters for both.

\subsection{Scheduling}\label{subsec:scheduling}

If geometrically nonlocal gates are allowed, and fresh ancilla qubits are always available on demand, then we can schedule the CZ gates so that no data qubit or ancilla qubit is idle during any time step. We assume that a preparation, measurement, or CZ gate can be performed in a single time step.  Suppose, for example, that we use the gadget design in \autoref{fig:CNOT-gadget}b, where the outgoing control data block participates in both $\MeasZZ^L$ and $\MeasZZZ^L$, and let us also assume that the [12] cat states have length $2m$ while the [234] cat states have length $3m$, matching the weight of the measured operators. We label the $r$ ancilla registers used to measure a row operator $r$ times by $a\in\{1,2,3,\dots,r\}$ for the $ZZ^L$ measurement and by $b\in\{1,2,3,\dots,r\}$ for the $ZZZ^L$ measurement. Suppose that $\PrepPlus^L[2]$ is completed in time step -1 and that $\PrepPlus^L[3]$ is completed in time step 0. Then the CZ gates are scheduled as follows: The $[12]$ ancilla $a$ interacts with block 2 in time step $a-1$ and with block 1 in time step $a$, while the $[234]$ ancilla $b$ interacts with blocks 2, 3, and 4 in time state $b$. Hence processing of the incoming blocks (1 and 4) begins in time step $1$, and processing of the outgoing blocks (2 and 3) ends in time step $r$. 

The ``preparation'' of the [234] ancilla $b$ begins in time step $b+1$, but the ``preparation'' of the [12] ancilla $a$ is ``staggered.'' The syndrome measurement for the first $m$ qubits in ancilla $a$ (those that interact with block 1) begins in time step $a + 1$, while the syndrome measurement for the last $m$ qubits (those that interact with block 2) begins one step earlier (step $a$). Despite this one time step delay, no qubits are ever idle --- the ancilla qubit used to measure syndrome bit $Z[m]Z[m+1]$ in cat state $a$  interacts with cat state qubit $m+1$ in time step $a$ and interacts with cat state qubit $m$ in time step $a+1$. Meanwhile, the ancilla qubit used to measure the syndrome bit $Z[1]Z[2m]$ in cat state $a$ interacts with cat state qubit $2m$ in time step $a+1$ and with cat state qubit 1 in time step $a+2$, {\em etc.} 


If instead we use the CNOT gadget design in \autoref{fig:CNOT-gadget}a, where the incoming control data block participates in both $\MeasZZ^L$ and $\MeasZZZ^L$, the processing of the data blocks is not perfectly synchronized. If $\MeasZZ^L$ is performed first, then the incoming and outgoing control blocks are one time step ahead of the incoming and outgoing target blocks, while if $\MeasZZZ^L$ is performed first it is the other way around. Thus if the incoming control block is either one step ahead or one step behind the incoming target block, we can perform the CNOT without leaving any qubits idle, while maintaining the one-step lag between the blocks. But if the two incoming blocks are synchronized, then using the gadget in \autoref{fig:CNOT-gadget}a one block or the other would have to wait for one time step before the processing of the block begins. 

\subsection{Injection by teleportation}

To complete a universal set of protected gates, we will use the magic state distillation procedure introduced by Bravyi and Kitaev \cite{Bravyi08}. First we inject non-Clifford states into the code block using teleportation, then we improve the fidelity of these encoded states via distillation. The distillation protocol is discussed further in Sec.~\ref{sec:distillation}; here we briefly describe the state injection procedure, with further details postponed until Sec.~\ref{sec:injection}.

The state injection makes use of the ``one-bit teleportation'' circuit depicted in \autoref{fig:teleportation}. To inject a state into the Bacon-Shor block, first the state $|\psi\rangle$ is prepared using a single physical qubit, while a logical block is prepared in the state $|+\rangle^L$. Then the weight-$(m+1)$ operator $Z^LZ$ is measured on the single qubit and a single row of the block, and finally the single qubit is measured in the $X$ basis. This procedure prepares the block in the state $|\psi\rangle^L$, up to a logical Pauli operator known from the measurement outcomes. 

There are other possible injection procedures, but this one is particularly simple. A single fault in the circuit can cause failure.

\section{Effective error strength}

The CNOT gadget may produce a logical error if any one of its four measurements deviates from the ideal case. To estimate the failure probability for the CNOT gadget, we must enumerate the ways in which these measurements might fail.

One potential source of trouble is the preparation of cat states and the logical $|+\rangle^L$ states. Recall that in these preparation gadgets a syndrome is measured repeatedly, and that the ``winning'' (\emph{i.e.}, most frequently occurring) syndrome measurement result is used to update the Pauli frame. This winning syndrome might differ from the syndrome that would have been inferred in an ideal gadget because of repeated $Z$ errors that induce errors in the syndrome measurement, because of $X$ errors that cause the syndrome to evolve between rounds, or because of some combination of the two. We say that the preparation gadget ``succeeds'' if the winning syndrome matches the syndrome that could have been obtained by an ideal measurement at some stage during the multiple rounds of syndrome measurement, and if furthermore fewer than half of the cat state qubits are afflicted by $X$ errors during the preparation process. Otherwise the preparation gadget ``fails.''

Failure of a preparation gadget can be catastrophic, because the inferred Pauli frame may differ from the actual Pauli frame by a high-weight $X$-type error. If it occurs in a cat state, this high-weight $X$ error could propagate to the data, causing a logical $Z^L$ error.  If it occurs in the preparation of a logical block, the error could flip the value of a subsequent $Z$-type logical measurement. 

On the other hand, even if a cat state preparation succeeds, the winning syndrome might not coincide perfectly with the ideal syndrome that applies to the cat state at the time it interacts with the data. The syndrome may drift due to $X$ errors that accumulate during the many rounds of syndrome measurement. However, the inferred Pauli frame differs from the ideal Pauli frame by a number of $X$ errors that is no larger than the number of non-diagonal faults that occur inside the preparation gadget. Though the inferred Pauli frame may not be exactly right, it is not catastrophically wrong unless there are many $X$ errors in the circuit. 

Aside from errors in the measured cat state syndrome, a $\MeasZZ^L$ or $\MeasZZZ^L$ could be faulty because of $Z$ errors acting on the cat state qubits, which flip the outcomes of  row measurements. Because each row measurement is repeated $r$ times, multiple errors are required for a row measurement to fail. Alternatively, an $X$ error acting on a data qubit could flip the outcome of a row measurement. Because a majority vote is performed on $n$ row measurement outcome, again multiple errors are required for the logical $Z$-type measurement to fail. An $\MeasX^L$ could be faulty because of $Z$ errors on data qubits, which flip outcomes of subsequent $X$ measurements on those qubits. Because a majority vote is performed on the $m$ column parities, multiple errors are required for the $\MeasX^L$ to fail.

In estimating the failure probability for $\MeasX^L$, $\MeasZZ^L$, or $\MeasZZZ^L$, we need to take into account possible faults in circuit locations preceding the CNOT gadget, which might cause these measurements to deviate from their ideal outcomes. In our fault-tolerant scheme, a logical CNOT gate is either preceded by teleportation into the code block or by another CNOT gate; in the worst case, each input block to the CNOT gadget is an output block from an immediately preceding CNOT gadget. Thus, in the preceding gadget, each block is prepared in the state $|+\rangle^L$, and then subjected to either $\MeasZZ^L$ and $\MeasZZZ^L$ (if it is the control block of the CNOT gadget) or just $\MeasZZZ^L$ (if it is the target block of the CNOT gadget). 

\subsection{Measurement failure}

We will assume that failure of any of its preparation gadgets will cause the CNOT to fail, and we denote by $\Perr^*$ the failure probability for a logical measurement due to a cause other than failure of a preparation gadget. Thus, the failure probability for the CNOT can be bounded as
%
%
\begin{widetext}
\begin{align}\label{eq:CNOT-L-error}
	\Perr(\mbox{CNOT$^L$}) ~ \leq & ~\Perr^*(\MeasZZ^L) + \Perr^*(\MeasZZZ^L) + 2 \Perr^*(\MeasX^L)  \nonumber\\ 
	& + 4 \Perr(\PrepPlus) + \Perr(\PrepZZCat) + \Perr(\PrepZZZCat) \nonumber\\
	& +  2\left( \Perr(\PrepZZCat) +\Perr(\PrepZZZCat) \right) .
\end{align}
\end{widetext}
The last term accounts for preparation errors that may have occurred in a $\MeasZZ^L$ or $\MeasZZZ^L$ immediately preceding the current CNOT gadget, acting on either one of the two input blocks to the current gadget. In the worst case, both incoming blocks were control blocks in the preceding CNOT gadgets, and were therefore subjected to both $\MeasZZ^L$ and $\MeasZZZ^L$. The coefficient of $\Perr(\PrepPlus)$ is 4 rather than 2 for a similar reason --- the logical measurement outcomes in the current CNOT gadget may differ from ideal outcomes due to errors in the preparation of logical blocks either in the current gadget or in one of the two preceding gadgets. 


Before considering the probability of a preparation failure, let us estimate $\Perr^*$ for each of the three types of logical measurements. For an $\MeasX^L$ measurement to fail, more than half of the $m$ length-$n$ columns must each have at least one $Z$ error.  In the worst case, each qubit in the block participates in $3r + 2r_+ + 2$ circuit locations: a preparation, a measurement, $2r_+$ CZ gates contained in the preparation of the logical $|+\rangle^L$ state, $r$ CZ gates in the logical $Z$ measurement in the current gadget, and $2r$ CZ gates in two logical $Z$ measurements in the preceding gadget. Each of these locations could be faulty with probability $\varepsilon + \varepsilon'$, since either a diagonal of non-diagonal fault could produce a phase error. 

For the moment, let us assume for simplicity that the cat state has length $2m$ for the $ZZ^L$ measurement and length $3m$ for the $ZZZ^L$ measurement, so that no ancilla qubit interacts with more than one data qubit. (We will consider the case where the cat state is shorter in Sec.~\ref{subsec:shorter}.) Because the preparation of the cat state actually occurs \emph{after} the ancilla interacts with the data, we do {\em not} need to worry about $X$ errors arising in the cat state preparation propagating directly to the data. We \emph{do} need to worry that the Pauli frame inferred from the ``winning'' syndrome after $r'$ rounds of syndrome measurement actually differs from the ideal Pauli frame. However, assuming that the syndrome is decoded correctly (as it will be if the preparation succeeds), a Pauli frame $X$ error afflicts a qubit only if at least one $X$ error acted on that qubit during the multi-round syndrome measurement. The probability of such a Pauli-frame $X$ error on any qubit in the cat state is therefore bounded above by $2r'\varepsilon'$, since two CZ gates act on each qubit in each round of syndrome measurement, and the syndrome measurement is repeated $r'$ times in the cat state preparation. Because each row measurement is repeated $r$ times, the total probability that a row measurement fails due to a Pauli-frame $X$ error in a cat state is bounded above by $2rr'\eps'$. Furthermore, in the worst case there are three logical $Z^L$ measurements between the preparation and measurement of the block, in each of which a cat state interacts with the data.

Combining together the probability per qubit of a $Z$ error caused directly by a fault with the probability of a $Z$ error that propagates from the cat state, we conclude that the error probabilities for the $X^L$ measurements can be bounded as
%
\begin{widetext}
\begin{equation} \label{eq:X-meas-error-prob}
	\Perr^*(\MeasX^L) \leq  \binom{m}{\frac{m+1}{2}} \left[ n(2 r_{+} + 3r + 2) (\eps+ \eps') + 6nrr'\eps'\right]^{(m+1)/2}.
\end{equation}
\end{widetext}
%
%
As explained in Sec.~\ref{subsec:scheduling}, there are no time steps in which ``resting'' qubits are subject to ``storage errors.'' 
%

For the $\MeasZZ^L$ to fail, more than half of the $n$ weight-$2m$ row parity measurements must have errors. An $X$ error acting on any one of the $2m$ qubits in the same row of the two blocks, either before or during the measurement, could flip the parity of the row. Before the $\MeasZZ^L$ is completed, each qubit participates in $2r_+$ CZ gates during the $\PrepPlus$ operation, $r$ CZ gates during the $\MeasZZ^L$, and in the worst case another $2r$ CZ gates during the $\MeasZZZ^L$ and $\MeasZZ^L$ in the preceding CNOT gadget. Thus the probability per qubit of a non-diagonal error is $(2r_+ + 3r)\eps'$. There might also be an $X$-type Pauli frame error in the $|+\rangle^L$ preparation, but we do not have to count that separately; if the preparation ``succeeds,'' then a Pauli-frame $X$ error afflicts a qubit in the block only if a non-diagonal fault damaged that same qubit during the preparation circuit. 

In the absence of non-diagonal faults, the row parity measurement might still fail because at least one $Z$ error acts on the cat state in each of $(r+1)/2$ of the repeated measurements. If the cat state has length $p$, then in each repetition of the measurement there are $2m + 2p + 2r'p$ circuit locations ($2m$ CZ gates coupling the ancilla to the data, plus $p$ qubit preparations, $p$ qubit measurements and $2pr'$ CZ gates for the cat state preparation). We conclude that the failure probability can be bounded as
\begin{widetext}
\begin{equation} \label{eq:ZZ-error-prob}
	\Perr^*(\MeasZZ^L) \leq \binom{n}{\frac{n+1}{2}} \left[ 2m(2 r_{+} + 3r) \eps' + \binom{r}{\frac{r+1}{2}}[(2m + 2p + 2p r')(\eps+\eps')]^{(r+1)/2} \right]^{(n+1)/2}.
\end{equation}

\noindent
Similarly,

\begin{equation} \label{eq:ZZZ-error-prob}
	\Perr^*(\MeasZZZ^L) \leq \binom{n}{\frac{n+1}{2}} \left[ 3m(2 r_{+} + 4r) \eps' + \binom{r}{\frac{r+1}{2}}[(3m + 2p + 2p r')(\eps + \eps')]^{(r+1)/2} \right]^{(n+1)/2} .
\end{equation}
\end{widetext}
(The $\MeasZZZ^L$ is preceded by a $\MeasZZ^L$ acting on one of its input blocks inside the current CNOT gadget, plus additional $Z$-type logical measurements in the preceding CNOT gadgets acting on another one of its input blocks.)

\subsection{Cat state preparation failure}
Now it remains to estimate the probability of failure for the cat state and the logical state $|+\rangle^L$. Recall that the preparation fails if the syndrome measurement is faulty in every round that yields the ``winning'' syndrome outcome. The faults can be either non-diagonal or diagonal, but there must be at least two faults; otherwise only one syndrome bit would be wrong, the result would fail the parity test, and the syndrome would be rejected. (A round with a single non-diagonal fault might be accepted if the fault flips a syndrome bit while also simultaneously applying an $X$ error to one of the cat state qubits, which flips a second syndrome bit measured in the following step. However, in this case the measured syndrome matches the actual cat state syndrome at the end of the round, and therefore is not counted as a failure.)

Suppose that the winning syndrome occurs $t$ times, and that there are $u$ additional rounds that each contain at least one fault (some of these rounds might be rejected). Suppose that non-diagonal faults occur in $s$ of the $r'$ rounds of syndrome measurement; these faults can alter the syndrome.  There are $r' -t -u$ rounds without any faults, and the number of distinct syndromes detected in these rounds is at most $s+1$. 

Now we can use the pigeonhole principle to obtain a lower bound on $t$, expressed in terms of $r'$, $u$, and $s$. There are at most $s+1$ valid syndromes that can occur in syndrome measurement rounds that have no faults. Combining these with the winning syndrome, there are at most $s+2$ possible syndromes that can occur in the $r'-u$ accepted rounds which either have no faults or produce the winning outcome. Of these $s+2$ syndromes, the winning syndrome must occur at least as many times as any other syndrome; hence
\begin{equation}\label{eq:t-bound}
	t \geq \left\lceil \frac{r'-u}{s+2} \right\rceil ,
\end{equation}
where $\lceil x \rceil$ denotes the smallest integer greater than or equal to $x$.

To bound $\Perr(\PrepCat)$, we sum over $s$ and $u$ in each cat state preparation step, estimating the number of possible fault histories using the upper bound Eq.~(\ref{eq:t-bound}) on $t$. In the first of the $t$ winning rounds, a particular winning syndrome is found, which differs in at least two bits from the actual syndrome in the beginning of that round. Then this same syndrome is found again in each of the remaining winning rounds. The sum over all possible winning syndromes, weighted by their probabilities, is bounded above by the probability that at least two measured syndrome bits are faulty in the first of the $t$ winning rounds. Each $ZZ$ measurement is performed using one $|+\rangle$ preparation, two $CZ$ gates, and one $X$ measurement; therefore the probability of error in the measurement of a single syndrome bit is bounded above by $4\eps + 2\eps'$, and the probability that at least two syndrome bits are in error is bounded above by $\binom{p}{2}(4\eps + 2\eps')^2$.

In each of the $s$ rounds that contain $X$ errors, we must sum over all the possible $X$-error patterns that can occur in that round. The sum over all $X$-error patterns, weighted by the probabilities, is bounded above by the probability that at least one nondiagonal fault occurs in that round. Since the round contains $2p$ $CZ$ gates, this probability is in turn bounded above by $2p\eps'$.

Once the $X$-error pattern has been chosen in each round that contains $X$ errors, we know the actual syndrome at the beginning of each round. And once the winning syndrome is chosen in the first winning round, we know which syndrome bits must have errors in each of the remaining $t-1$ winning rounds. If the cat state preparation fails, then by definition at least two syndrome bits have errors in each of these rounds; hence each winning round after the first has a probability weight bounded above by $(4\eps + 2\eps')^2$.


Taking into account that the rounds with non-diagonal faults can be chosen in at most $\binom{r'}{s}$ ways, and enumerating the ways to choose which $t$ rounds produce the winning syndrome and which $u$ additional rounds have faults, we obtain
\begin{widetext}
\begin{equation} \label{eq:cat-prep-bound}
	\Perr(\PrepCat) \leq n r \sum_{s=0}^{r'} \sum_{u=0}^{r'} \binom{r'}{s} \binom{r'}{u+t} \binom{u+t}{u} \binom{p}{2}(4\eps + 2\eps')^{2t} (4p\eps + 2p\eps')^u (2p \eps')^s ,
\end{equation}
and similarly
\begin{equation} \label{eq:plus-prep-bound}
	\Perr(\PrepPlus) \leq m \sum_{s=0}^{r_+} \sum_{u=0}^{r_+} \binom{r_+}{s} \binom{r_+}{u+t} \binom{u+t}{u} \binom{n}{2}(4\eps + 2\eps')^{2t} (4n\eps + 2n\eps')^u (2n \eps')^s ,
\end{equation}
\end{widetext}
where $t$ denotes $\lceil \frac{r'-u}{s+2} \rceil$. The prefactor $nr$ in Eq.~(\ref{eq:cat-prep-bound}) arises because we use $n$ length-$p$ cat states in each measurement, and each measurement is repeated $r$ times. The prefactor $m$ in Eq.~(\ref{eq:plus-prep-bound}) arises because the encoded state $|+\rangle^L$ is a product of $m$ length-$n$ cat states.



We should add another contribution to the failure probability, because if half of the qubits  (or more) have $X$ errors, we might decode the cat-state syndrome incorrectly. A syndrome of the repetition code points to two possible $X$ error patterns, one low weight and one high weight. We always assume the low-weight interpretation is correct, so if the high-weight interpretation is actually correct, then a Pauli-frame error $X^\otimes p$ is applied to the cat state (or worse if the cat state qubits are used multiple times). The additional contribution, then, is bounded by the probability that each of $\lceil p/2 \rceil$ qubits in the cat state are each hit by $X$ errors at least once in at least one of the cat state preparations. This upper bound is
\begin{equation}
nr \binom{p}{\lceil p/2\rceil}\left( 2r'\eps' \right)^{\lceil p/2\rceil}.
\end{equation}

The Kraus operator arising from a general non-diagonal fault could be a coherent linear combination of Pauli operators; therefore the pattern of $X$ errors acting on the cat state may be a coherent superposition of many possibility. For convenience and clarity, we assumed in the above discussion that non-diagonal errors are actually Pauli errors, so that $X$-error patterns can be assigned probabilities rather than amplitudes. But the argument also applies to more general non-diagonal faults.  --- by expanding each Kraus operator in terms of Pauli operators, and noting that all syndrome bits are measured in the $X$ basis, we can bound the probability for each syndrome measurement outcome in terms of $\eps$ and $\eps'$ as above.

\subsection{Shorter ancillas}\label{subsec:shorter}
Cat states are used in measurements of the weight-$2m$ logical operator $ZZ^L$ and the weight-$3m$ logical operator $ZZZ^L$. In the derivation of Eq.~(\ref{eq:X-meas-error-prob}) we assumed that the cat states have length $2m$ and $3m$ respectively, so that each of the cat state's qubits participates in just one CZ gate during the measurement. But we have found that the CNOT gadget may be more reliable if a shorter cat state is used instead. 

If the length of the cat state is at least $m$, then each cat-state qubit interacts with only one qubit in each encoded block. But if the length is $p < m$, then a single cat-state qubit may interact with as many as 
\begin{equation}
R = \lceil m/p \rceil
\end{equation} 
data qubits in a single block, and hence an $X$ error on that cat-state qubit may propagate to produce $Z$ errors in $R$ distinct columns of one block.
If $k$ of the cat state's $p$ qubits have $X$ errors, these could generate $Z$ errors in $kR$ columns of the block; thus an $X^L$ measurement on the block could fail if there are
\begin{equation}
L(k) = (m+1)/2 - R k 
\end{equation}
additional columns that each contain at least one $Z$ error. 

In each measurement, each data block interacts with the ancilla for $R$ consecutive time steps, where in each step at most $p$ CZ gates act on the data block. The repeated measurements can be staggered as described in \cite{Aliferis08}, to avoid time steps in which data qubits are idle.


We use the index $i\in\{1,2,3,\dots,p\}$ to label the positions of the $p$ qubits in the cat state, and design the measurement circuits so that the cat-state qubit $i$ interacts with the same set of data qubits in each one of the measured code blocks. Since the $ZZ^L$ measurement is performed on all $n$ rows of each block, and also repeated $r$ times, inside a $\MeasZZ^L$ there are at most $2nrR$ CZ gates  that act on a cat-state qubit at position $i$ (not counting CZ gates in the cat state preparation step), any of which could have a non-diagonal fault. Similarly, inside a $\MeasZZZ^L$ there are at most $3nrR$ CZ gates  that act on a cat-state qubit at position $i$. 

The $\MeasX^L$ inside a CNOT gadget is preceded by $\MeasZZZ^L$ or $\MeasZZ^L$ within the CNOT, and in the worst case by both $\MeasZZZ^L$ and $\MeasZZ^L$ in the preceding gadget, so at most there are $8nrR$ CZ gates in these measurements where a non-diagonal fault could disturb cat-state qubit $i$. 
An $X$ error acting on cat-state qubit $i$ might also occur because of an error in the winning syndrome in one of the cat-state preparation steps, which can occur only if a faulty $CZ$ gate acts on qubit $i$ at least once during the $r'$ repeated syndrome measurements. Noting that a damaged cat-state qubit in position $i$ could cause $Z$ errors in $R$ columns of the encoded block, and that $k$ damaged qubits in the length-$p$ cat state can be chosen in $\binom{p}{k}$ ways, we conclude that if length-$p$ cat states are used then Eq.~(\ref{eq:X-meas-error-prob}) should be replaced by

%
\begin{widetext}
\begin{equation}
\Perr^*(\MeasX^L) \leq  \sum_{k=0}^{k_{\text{max}}} \binom{p}{k}  \left[ 8 n rR \eps' + 6 n rr'\eps' \right]^k \binom{m}{L(k)} \left[ n (2 r_{+} + 3r +2) (\eps+\eps') \right]^{L(k)} ,
\end{equation}
\end{widetext}
where
\begin{align} 
	k_{\text{max}} &= \lceil (m+1)/2 R\rceil .
\end{align}


\section{Geometrically local circuits}
\label{sec:local}

In our analysis so far, we have assumed that the two-qubit CZ gates can be performed on any pair of qubits, with an error rate independent of the distance between the qubits. Now we will consider how the analysis is modified if CZ gates can be performed only on neighboring pairs of qubits.

There are a variety of possible architectures for fault-tolerant quantum computing using Bacon-Shor codes. To be concrete, we will consider an effectively one-dimensional arrangement, in which logical qubits are encoded in a ribbon of physical qubits with constant width in the vertical direction and length in the horizontal direction proportional to the total number of logical qubits. Each logical qubit lives in a ``bi-block'' --- a pair of Bacon-Shor code blocks, one storing the data processed by the computation, and the other used as an auxiliary block for teleporting gates. If desired, the data block can be shuffled back and forth between the left and right sides of the bi-block via a ``one-bit-teleportation'' circuit consisting of a $|+\rangle^L$ preparation, a $ZZ^L$ measurement, and an $X^L$ measurement, as in \autoref{fig:teleportation}.

\begin{figure}[ht]
 	\begin{center}
 	\includegraphics[width=0.45\textwidth]{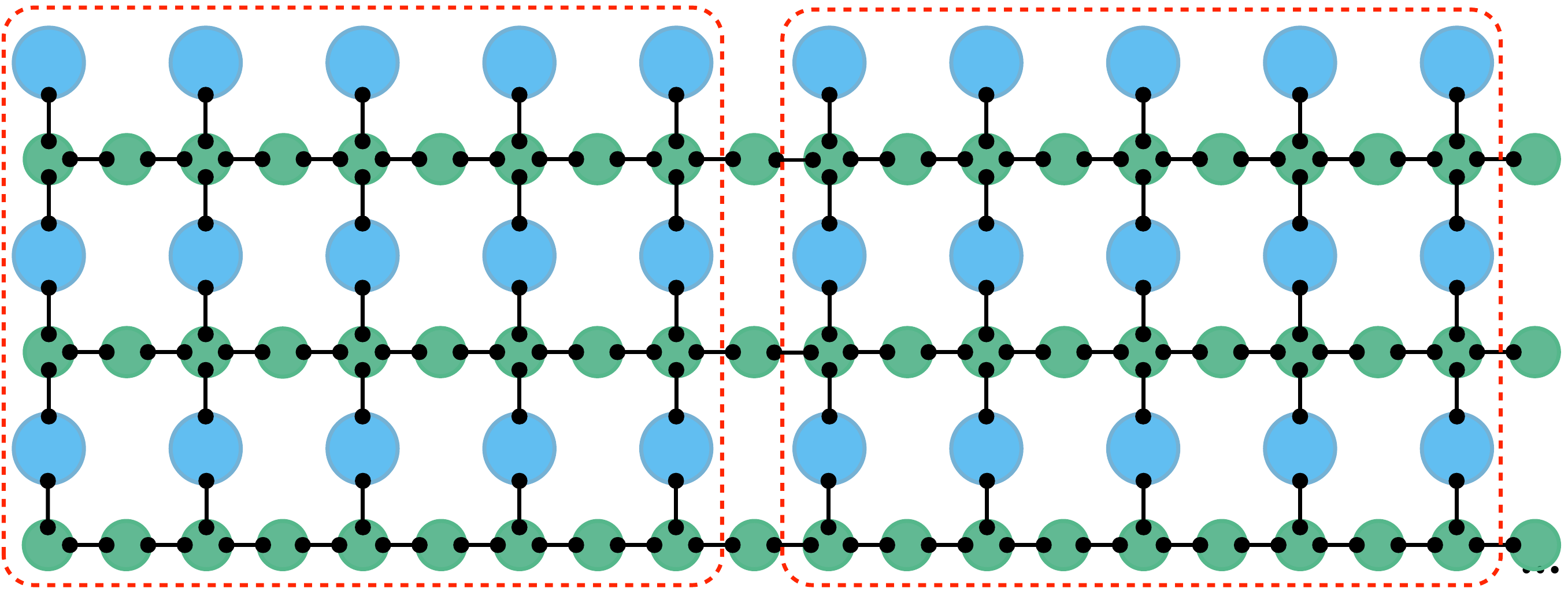}
 	\end{center}
	\caption{(Color online) Arrangement of Bacon-Shor qubits for geometrically local computation for $3 \times 5$ code block. Large blue circles indicate data qubits, small green circles ancilla qubits. Solid lines indicate locations between which CZ gates may be performed. Dotted lines separate qubits into code blocks. The ribbon may be extended arbitrarily far to the right and left with additional code blocks.}
	\label{fig:lattice-local}
\end{figure}

Each $n\times m$ logical block is interlaced with an $n\times (2m-1)$ array of ancilla qubits, as in \autoref{fig:lattice-local}. Of the $2m-1$ ancilla qubits in a row, $m-1$ are used to read out the outcome of a $ZZ$ measurement performed on the qubit's two horizontal neighbors, preparing an $m$-qubit cat state in the row. In addition, there is a column of $n$ ancilla qubits at the boundary between two adjacent blocks, allowing the cat states to be extended to length $2m$ or $3m$ as needed for the measurements $\MeasZZ^L$ and $\MeasZZZ^L$ in the CNOT gadget. In the $|+\rangle^L$ preparation gadget, $n-1$ ancilla qubits in a column are used to read out the outcome of a $ZZ$ measurement on the qubit's two vertical neighbors, preparing an $n$-qubit cat state in a column.

A CNOT gate is performed on a neighboring pair of logical bi-blocks using one of the variants of the CNOT gadget depicted in \autoref{fig:CNOT-gadget}. Let us denote the bi-block on the left as bi-block 1 and the bi-block on the right as bi-block 2. If the data in bi-block 1 is on the right side of the bi-block and the data in bi-block 2 is on the left side,
then we may use the CNOT gadget design in \autoref{fig:CNOT-gadget}$a$ to perform a CNOT gate in either direction ({\em i.e.} with either logical qubit as the control and the other as the target) using cat states of length $2m$ and $3m$, with the output data from the gate appearing on the left side of bi-block 1 and the right side of bi-block 2. Otherwise, we may use one of the alternative gadget designs discussed in \autoref{fig:CNOT-gadget} to deal with the cases where the data in bi-block 1 is on the left side and/or the data in bi-block 2 is on the right side. Whichever design we choose, the CNOT gadget flips the data from one side to the other in both bi-blocks, and only the $\MeasZZZ^L$ measurement reaches across the boundary between the two bi-blocks.

Note that with this method for performing a logical CNOT gate using geometrically local gates there is no need to swap the positions of pairs of physical qubits. The CZ gate is the only two-qubit gate used at the physical level, so that the assumption of highly biased noise remains physically plausible. Bacon-Shor blocks can be swapped using circuits of logical CNOT gates.

To complete a universal gate set we also need to be able to inject noisy non-Clifford states into code blocks which can then be purified by state distillation. This state injection can be performed using one-bit teleportation --- the state $|\psi\rangle$ is prepared using a single physical qubit, while a logical block is prepared in the state $|+\rangle^L$, then the weight-$(m+1)$ operator $Z^LZ$ is measured on the single qubit and a single row of the block, and finally the single qubit is measured in the $X$ basis. This procedure prepares the block in the state $|\psi\rangle^L$, up to a logical Pauli operator known from the measurement outcomes. To perform this task in a bi-block, we prepare $|+\rangle^L$ on (say) the left side of the bi-block, and build a length-$(m+1)$ cat state using $m$ of the ancilla qubits that accompany a row of the block on the left side, and one additional ancilla qubit on the right side. The single-qubit state $|\psi\rangle$ is prepared using one physical qubit on the right side of the bi-block which is adjacent to the $(m+1)$st qubit of the cat state, allowing the measurement of $Z^LZ$ to be complete using local CZ gates.

If we wish to build a two-dimensional architecture, we may stack horizontal ribbons of logical qubits on top of one another. To perform CNOT gates on pairs of logical bi-blocks that are stacked vertically, we need to be able to perform $\MeasZZZ^L$ using cat states shared by vertically stacked blocks. For this purpose we can use ancilla qubits in neighboring bi-blocks to establish a ``channel'' linking the cat states neighboring corresponding rows of the vertically stacked blocks. 

We note that in the case of unbiased noise, a two-dimensional architecture might be realized more simply, using symmetric Bacon-Shor codes. In that case we can build a CNOT gate using either the gadget in \autoref{fig:CNOT-gadget} or its Hadamard dual, with $X$ and $Z$ interchanged. The primal gadget uses horizontal cat states to execute CNOT gates on horizontally neighboring blocks, while the dual gadget uses vertical (dual) cat states to execute CNOT gates on vertically neighboring blocks. 

For the geometrically local case, our analysis needs to be modified in several ways. As already indicated, we will not be able to choose the length $p$ of the cat state to optimize the gadget. Rather, the locality constraint requires that the length of the cat state match the weight of the measured operator ($p=2m$ for $\MeasZZ^L$ and $p=3m$ for $\MeasZZZ^L$). Another change is that we must now consider the consequences of storage errors acting on idle qubits. Previously, we assumed that the off-line preparation of cat states can be scheduled so that the cat states are always available immediately when needed. But now there is only one set of $m$ ancilla qubits to accompany each row of a data block. Each time a cat state is measured we must prepare a new one, and the preparation involves a syndrome measurement repeated $r'$ times. In each round of syndrome measurement, all qubits except the ones at the ends of the cat state participate in two $ZZ$ measurements, and because the same ancilla qubits are used for two different $ZZ$ measurements, four time steps are required for each one (a preparation, two CZ gates, and an $X$ measurement).  Therefore, the data waits for $8r'$ time steps before the cat state is ready to interact with the data. While the data qubits wait for the cat state to be ready, they may be subject to storage errors. We must estimate how these storage errors contribute to the failure probability of the logical measurements in the CNOT gadget. 

We will denote by $\eps_s$ the probability per time step of a diagonal storage error and by $\eps_s'$ the probability per time step of a non-diagonal storage error. How many storage steps are included depends on the gadget design, which in turn depends on whether the data is on the left or right side of the incoming bi-blocks. In the worst case, the measurement $\MeasX^L$ is performed on a block that undergoes four logical $Z$-type measurements between its preparation and measurement, two in the current gadget plus another two in the immediately preceding gadget, and for each of these measurements the data waits for $8r'$ steps as the cat state is prepared. If each $Z$-type logical measurement is repeated $r$ times, then the total contribution to the $Z$ error probability per qubit due to the identity gates in the circuit is $32rr'(\eps_s+ \eps'_s)$. Therefore, Eq.~(\ref{eq:X-meas-error-prob}) is modified to become
\begin{widetext}
\begin{equation}
 \label{eq:X-meas-error-prob-local}
	\Perr^*(\MeasX^L) \leq  \binom{m}{\frac{m+1}{2}} \left[ n(2 r_{+} + 3r + 2) (\eps+ \eps') + 32nrr'(\eps_s+ \eps'_s)+ 8nrr'\eps'\right]^{(m+1)/2}.
\end{equation}
\end{widetext}

Furthermore, a non-diagonal storage fault on any of the $2m$ data qubits can flip the outcome of $\MeasZZ^L$. This fault can occur in any of the $8r'$ time steps while a cat state is prepared in either the $\MeasZZ^L$ or one of the preceding measurement. Similarly, a non-diagonal storage fault on any of the $3m$ data qubits can flip the outcome of $\MeasZZZ^L$, where his fault can occur in any of the $8r'$ time steps while the cat state is prepared in either the $\MeasZZZ^L$ or one of the preceding measurements. Thus Eq.~(\ref{eq:ZZ-error-prob},\ref{eq:ZZZ-error-prob}) are replaced by
{\footnotesize
\begin{widetext}
\begin{align} \label{eq:Z-meas-error-prob-local}
	\Perr^*(\MeasZZ^L) \leq & \binom{n}{\frac{n+1}{2}} \left[ 2m(2 r_{+} + 3r) \eps' + 2m(24rr'\eps_s')+ \binom{r}{\frac{r+1}{2}}[(2m + 2p + 2p r')(\eps+\eps')]^{(r+1)/2} \right]^{(n+1)/2},\nonumber\\
	\Perr^*(\MeasZZZ^L) \leq &\binom{n}{\frac{n+1}{2}} \left[ 3m(2 r_{+} + 4r) \eps' + 3m(32rr'\eps_s') + \binom{r}{\frac{r+1}{2}}[(3m + 2p + 2p r')(\eps + \eps')]^{(r+1)/2} \right]^{(n+1)/2} .
\end{align}
\end{widetext}
}
Note that the number of storage locations in the circuit might be reduced by adding additional ancilla sites, so that more measurements can be performed in parallel, and/or by combining together measurement locations with immediately following preparation locations. 

Geometric locality also requires that we remove the redundant cat state measurements that ``wrap around'' the code block;  Thus there is no syndrome parity check; any syndrome outcome is accepted, and a single diagonal fault can generate a syndrome error. There are also storage locations where the cat state qubits wait while ancilla qubits are prepared and measured. Because of these changes, Eq.~(\ref{eq:cat-prep-bound},\ref{eq:plus-prep-bound}) are replaced by
\begin{widetext}
\begin{equation} 
	\Perr(\PrepCat) \leq n r \sum_{s=0}^{r'} \sum_{u=0}^{r'} \binom{r'}{s} \binom{r'}{u+t} \binom{u+t}{u}  \left[4p\eps + 2p\eps'+ 4p(\eps_s + \eps_s')\right]^{t+u} \left(2p \eps' + 4p\eps_s'\right)^s 
\end{equation}
and
\begin{equation} 
	\Perr(\PrepPlus) \leq m \sum_{s=0}^{r_+} \sum_{u=0}^{r_+} \binom{r_+}{s} \binom{r_+}{u+t} \binom{u+t}{u} \left[4n\eps + 2n\eps'+4n(\eps_s+\eps_s')\right]^{t+u} (2n \eps'+4n\eps_s')^s .
\end{equation}
\end{widetext}


\section{Results for logical CSS gates}

\subsection{Unrestricted gates}

\begin{figure*}[ht]
 	\begin{center}
 	\includegraphics[width=5in]{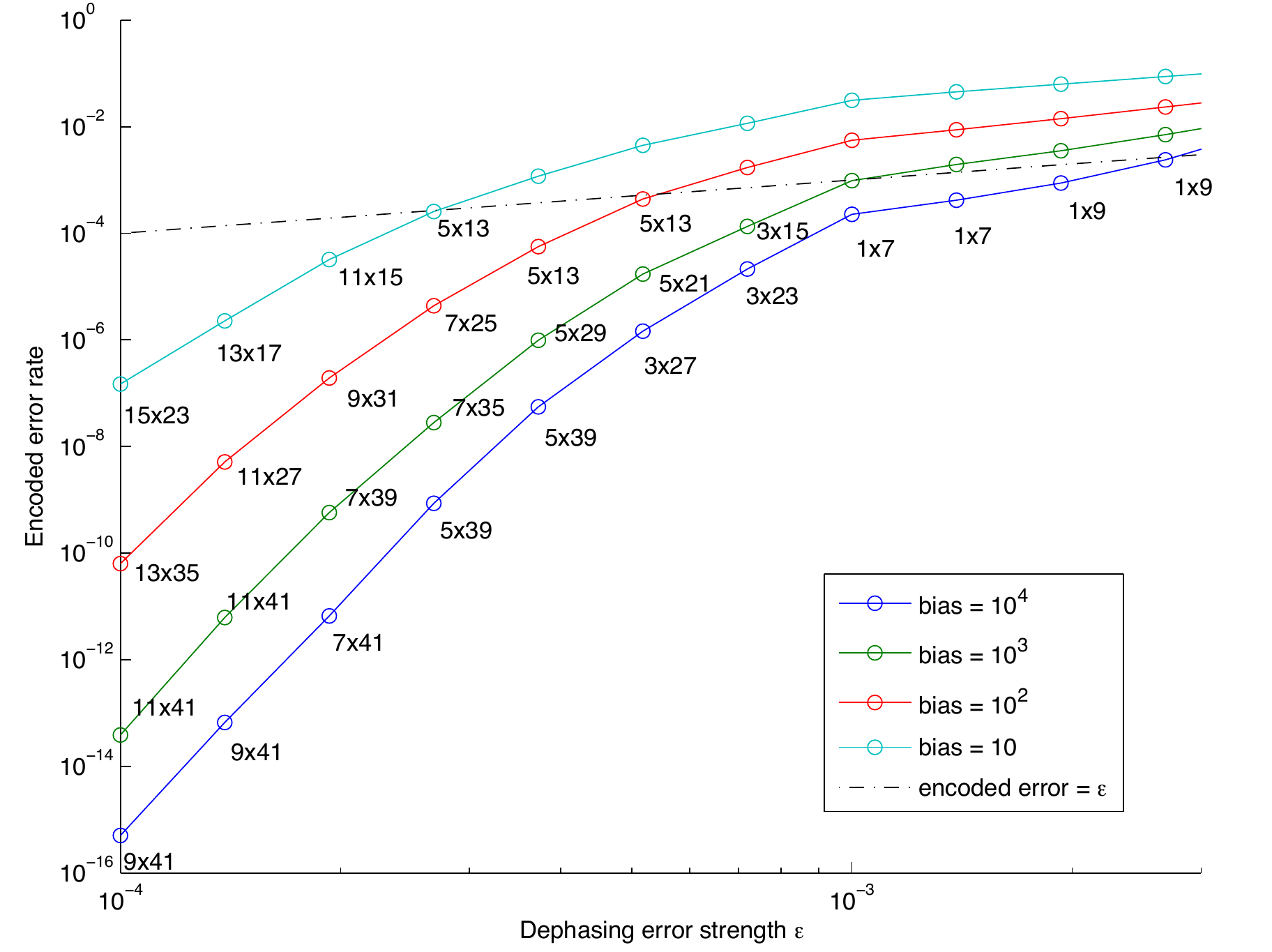}
 	\end{center}
	\caption{(Color online) Optimal CNOT logical error rate versus physical error rate for various values of the bias. Labels represent the $n \times m$ dimensions of the Bacon-Shor code block. For blocks with a single row, one can use the code studied in \cite{Aliferis08} which is a special case of the asymmetric Bacon-Shor code.}
	\label{fig:encoded-err-vs-err}
\end{figure*}

With our analytic upper bound on the effective error strength $\eps^{(1)}$ of the CNOT gadget, we can find the optimum choice of the code block size $n \times m$ as well as the four CNOT parameters $r$, $r'$, $r_+$ and $p$, for any choice of the error strengths $\eps$ and $\eps'$. We did this by brute force search over the parameter space. \autoref{fig:encoded-err-vs-err} plots the results of this optimization for five choices of the bias $R=\eps/\eps'$.

These results were obtained by optimizing the size of a Bacon-Shor code block which is used by itself, rather than as part of a more complex concatenated code. Alternatively, we may consider using our Bacon-Shor gadgets at the bottom level of a concatenated coding scheme. In particular, we may estimate the accuracy threshold for biased noise achieved by such concatenated codes, as was done in Ref. \cite{Aliferis08} for the special case of the $n=1$ Bacon-Shor code ({\em i.e.}, the repetition code). We have extended the analysis of the accuracy threshold performed in \cite{Aliferis08} to more general Bacon-Shor codes. However, for bias above $10^3$ we found no improvement over the threshold estimate in \cite{Aliferis08}, because the $n=1$ code turns out to provide the best threshold value. For values of the bias between 1 and $10^3$, the optimal value of $n$ turns out to be greater than 1, and for that range of the bias we found modest improvements in the accuracy threshold estimate compared to \cite{Aliferis08}.

\begin{figure*}[ht]
 	\begin{center}
 	\includegraphics[width=5in]{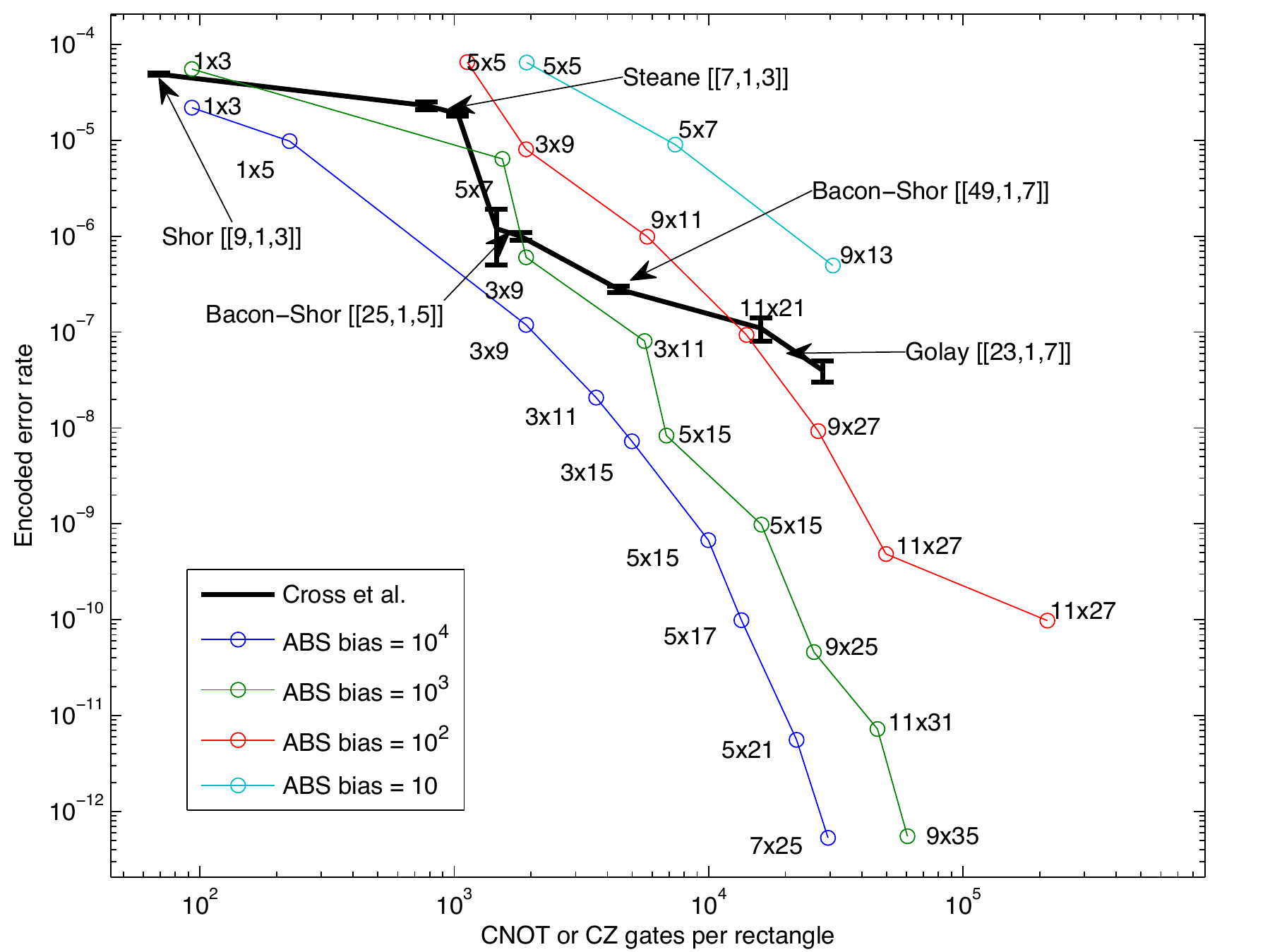}
 	\end{center}
	\caption{(Color online) Optimal CNOT logical error rate versus required number of physical gates for $n\times m$ asymmetric Bacon-Shor codes, for a physical error rate $\eps=10^{-4}$ at various values of the bias, plotted alongside results for codes studied in \cite{Cross07} with no bias.}
	\label{fig:encoded-err-vs-gates}
\end{figure*}

Using asymmetric Bacon-Shor codes, we can exploit the noise bias to improve the number of physical gates needed to construct a fault-tolerant CNOT gate with a specified logical error rate. \autoref{fig:encoded-err-vs-gates} shows this overhead factor and also indicates the dimensions of the optimal code block. This plot includes a comparison with the performance of concatenated codes surveyed in  \cite{Cross07} for the case of unbiased noise.  For highly biased noise, asymmetric Bacon-Shor codes achieve a much lower logical error rate with a smaller number of physical two-qubit gates, compared to the performance of these previous constructions for unbiased noise.

\subsection{Geometrically local gates}

\begin{figure}[ht]
 	\begin{center}
 	\includegraphics[width=0.45\textwidth]{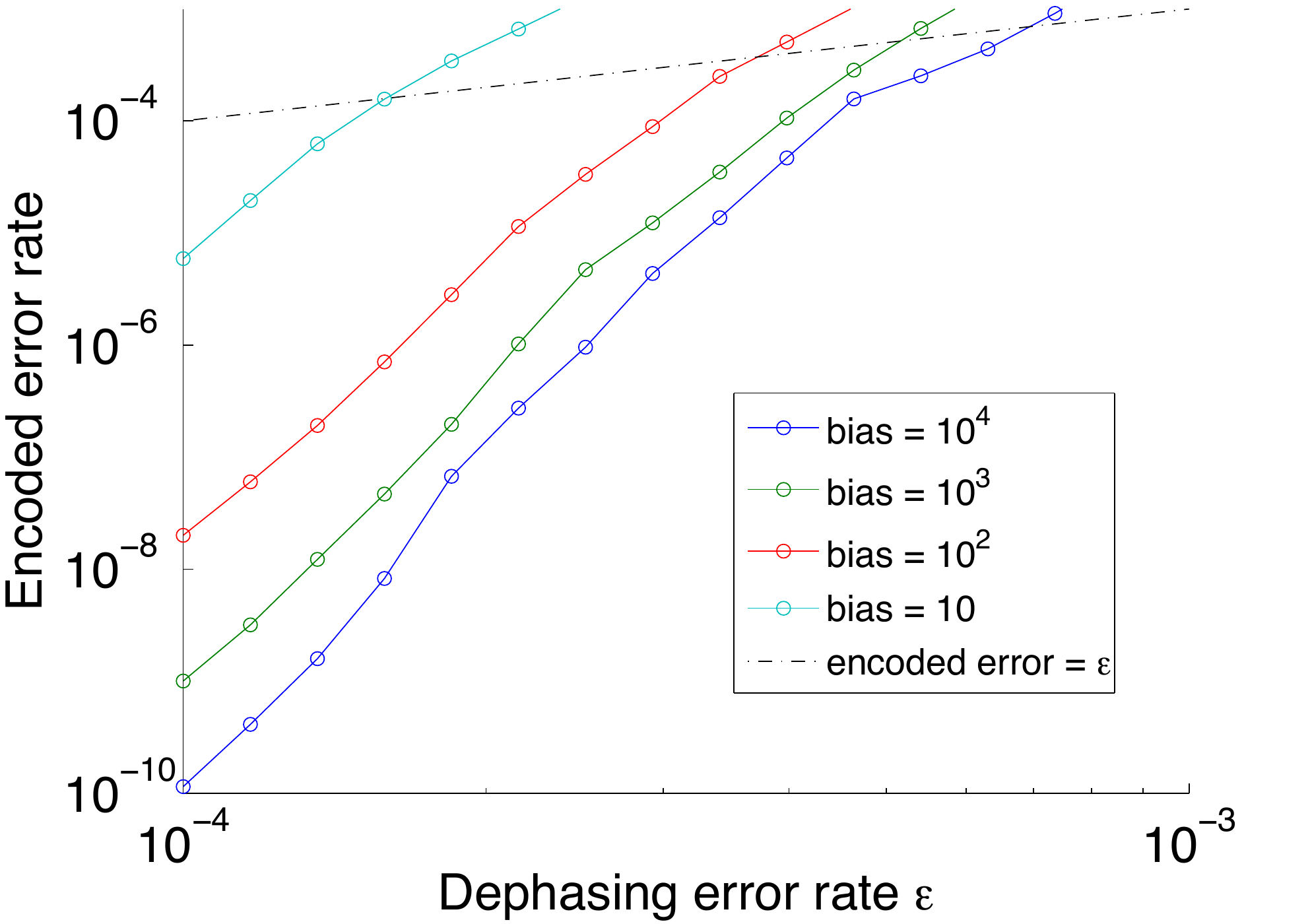}
 	\end{center}
	\caption{(Color online) Optimal CNOT logical error rate versus physical error rate for various values of the bias, where two-qubit physical gates are assumed to be geometrically local.}
	\label{fig:encoded-err-local}
\end{figure}

\begin{figure}[ht]
 	\begin{center}
 	\includegraphics[width=0.45\textwidth]{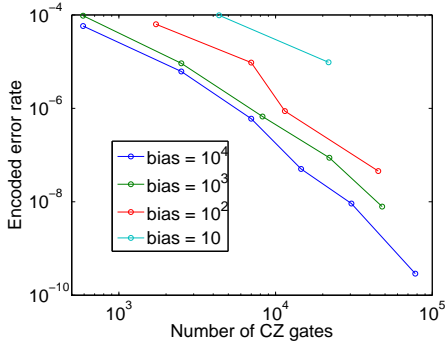}
 	\end{center}
	\caption{(Color online) Optimal CNOT logical error rate versus required number of physical gates for geometrically local asymmetric Bacon-Shor codes at various values of the bias, for a physical error rate $\eps=10^{-4}$}
	\label{fig:encoded-err-vs-gates-local}
\end{figure}

\begin{figure}[ht]
 	\begin{center}
 	\includegraphics[width=0.45\textwidth]{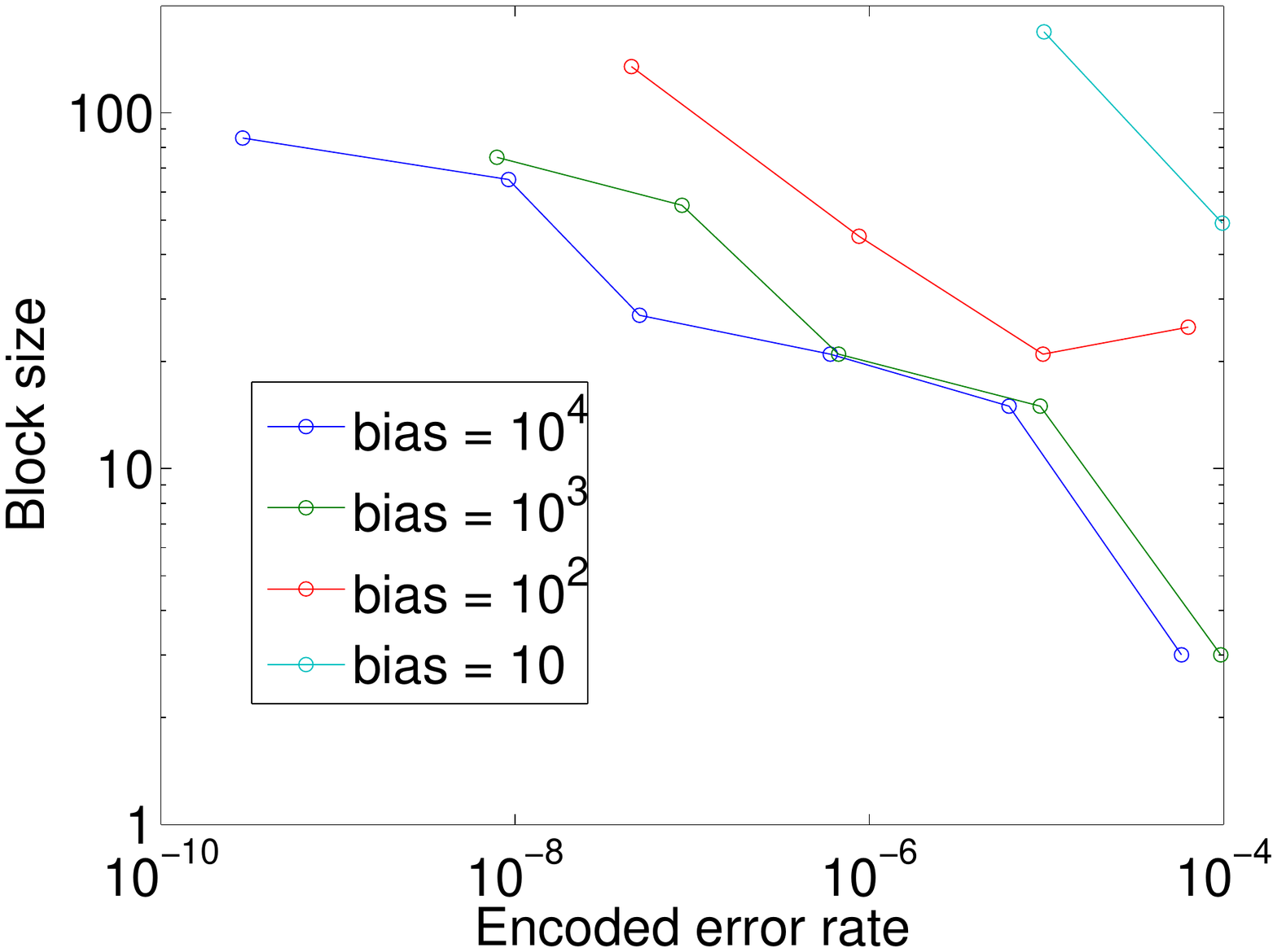}
 	\end{center}
	\caption{(Color online) Bacon-Shor block size vs optimal (in number of gates) CNOT logical error rate for geometrically local asymmetric Bacon-Shor codes at various values of the bias, for a physical error rate $\eps=10^{-4}$.}
	\label{fig:encoded-err-vs-block-size-local}
\end{figure}

Optimizing the parameters for the case of geometrically local gates, assuming the storage error rate is negligible, we obtain the results displayed in \autoref{fig:encoded-err-local} and \autoref{fig:encoded-err-vs-gates-local}. Enforcing locality significantly weakens the performance of our constructions --- at a bias of $10^4$ and dephasing error strength of $\eps = 10^{-4}$, the optimal effective error strength we can achieve increases from $10^{-20}$ to $10^{-9}$, while requiring roughly 8 times as many gates. Therefore, in the geometrically local case our estimated logical error rate for asymmetric Bacon-Shor codes and highly biased noise is roughly similar to what can be achieved using surface codes with a similar number of logical gates, while disregarding the bias \cite{raussendorf2007topological,fowler2009high}. The most important reason that nonlocal gates prove to be advantageous is that if geometry is ignored then the cat state used in (for example) the measurement $\MeasZZZ^L$ can be chosen to be much shorter than the length-$3m$ cat state used in our geometrically local construction. The number of physical qubits per block required to achieve the optimal logical error rate stays relatively small, as indicated in \autoref{fig:encoded-err-vs-block-size-local}.

Our upper bound on the logical error rate is likely to be far from optimal, and we also expect that the performance could be substantially improved by using a more sophisticated (but harder to analyze) method like perfect matching 
for decoding the cat state syndrome history. Suitable lattice deformations may also make it more feasible to reduce the length of cat states substantially, yielding further enhancements in performance for the case of geometrically local gates. Such improvements might make asymmetric Bacon-Shor codes more competitive relative to surface codes.

\subsection{Geometrically local gates and measurement bias}
\label{subsec:measurement-bias}

\begin{figure}[ht]
 	\begin{center}
 	\includegraphics[width=0.45\textwidth]{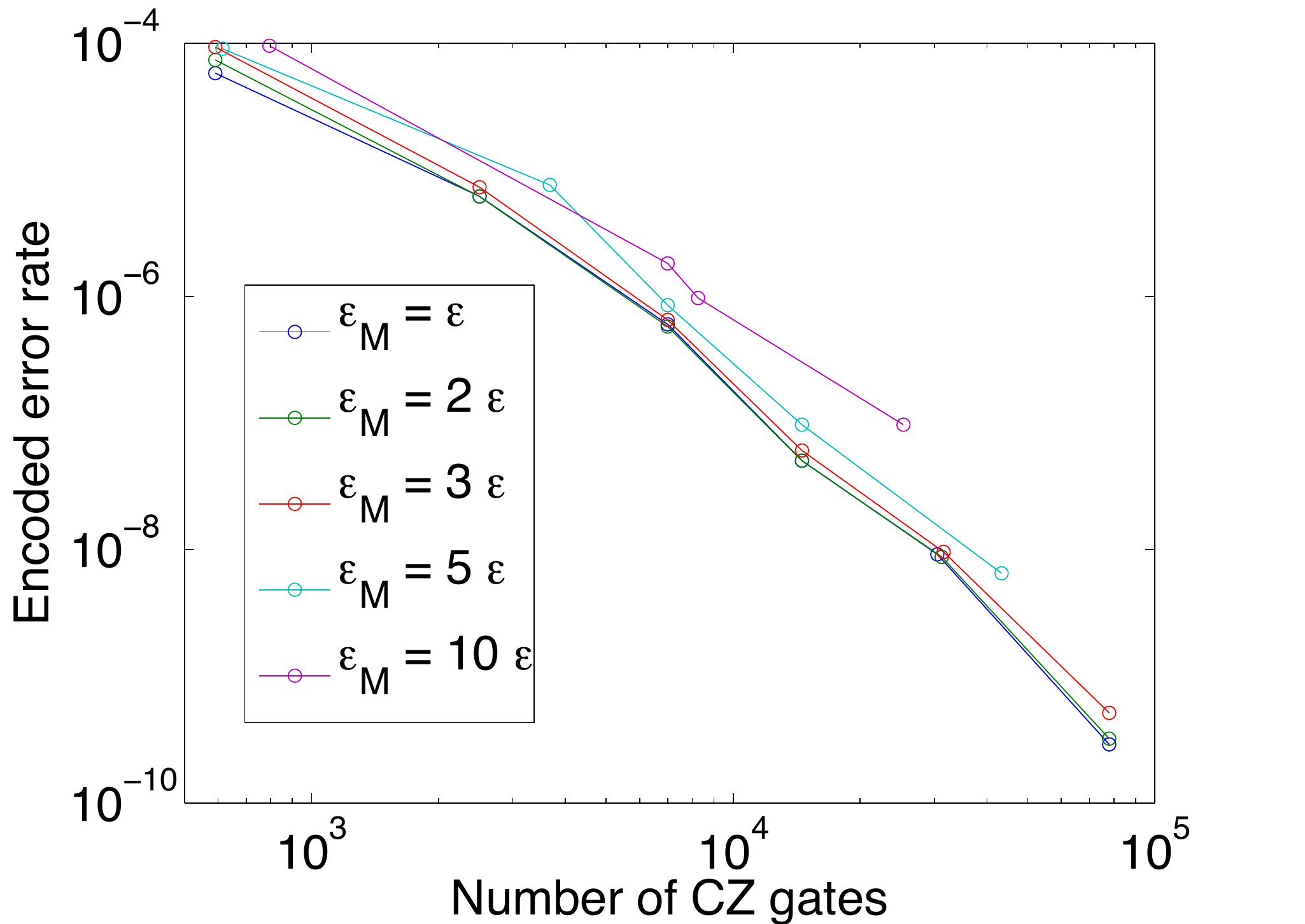}
 	\end{center}
	\caption{(Color online) Optimal CNOT logical error rate versus required number of gates for geometrically local asymmetric Bacon-Shor codes, assuming a physical error rate $\eps=10^{-4}$, bias $10^4$, and various rates of the measurement error rate $\eps_\mathcal{M}$.}
	\label{fig:encoded-err-vs-gates-meas-bias}
\end{figure}

In some experimental settings measurements are noisier than gates, in which case we say the noise has ``measurement bias.'' We may consider a noise model in which the dephasing error rate $\eps$ in diagonal gates exceeds the rate $\eps'$ for nondiagonal faults, while in addition the error rate $\eps_\mathcal{M}$ for single-qubit measurements exceeds $\eps$. (We also continue to assume that $\eps$ is the error rate for single-qubit preparations.) As shown in \autoref{fig:encoded-err-vs-gates-meas-bias}, our geometrically local asymmetric Bacon-Shor code gadgets are somewhat robust against increasing measurement bias, because our gadgets contain considerably more gates than measurements (or preparations). Relative to the case $\eps_\mathcal{M}=\eps$, the performance of the logical CNOT gate is not much affected as the measurement bias $\eps_\mathcal{M}/\eps$ rises to about 5. In contrast, surface-code gadgets, which contain a higher number of measurements relative to the number of gates, are more sensitive to measurement bias.

\section{State injection}\label{sec:injection}
Having analyzed the performance of our fault-tolerant CSS gates, we now turn to the state injection and distillation protocols needed to complete a universal set of fault-tolerant gates. An arbitrary single-qubit state $|\psi\rangle$ can be injected into the Bacon-Shor block using the ``one-bit teleportation'' circuit depicted in \autoref{fig:teleportation}. Here a logical qubit is prepared in the state $|+\rangle^L$, the two-qubit measurement $\mathcal{M}_{ZZ}^L$ is performed jointly on an unprotected qubit and the Bacon-Shor block using a length-$(m+1)$ cat state, and finally the $X$-basis measurement $\mathcal{M}_{X}$ is performed on the unprotected qubit. To estimate the error in the state injection step, we should consider all the ways in which the outcomes of the measurements $\mathcal{M}_{ZZ}^L$ and $\mathcal{M}_{X}$ might deviate from their ideal values.

In the state distillation circuit, this injection step is directly followed by a CNOT gate. Hence some sources of error in the teleportation circuit need not be attributed to the state injection step, as they are already included in our error estimate for the CNOT gate that follows. In particular, the possibility of failure in the preparation of $|+\rangle^L$ or a cat state is included in Eq.(\ref{eq:CNOT-L-error}), and conservatively at that, since we assumed there that the cat states had length $2m$ or $3m$ rather than $m+1$. Here we assume that the number of repetitions $r'$ of the cat state syndrome measurement, and the number of syndrome measurement repetitions $r_+$ in the  $|+\rangle^L$ state preparation, match the number of repetitions in the following CNOT gate. If so, we may bound the injection error probability $\Perr(\Inject)$ by

\begin{align}
	\Perr(\Inject) & \leq \eps_{\psi} + \Perr^*(\Meas_{Z Z}^L) + \Perr^*(\Meas_{X}),
\end{align}
where $\eps_\psi$ bounds the probability of an error in the preparation of the unprotected state $\ket{\psi}$ (we presume that this error is not necessarily diagonal in the $Z$ basis).

We assume that $\mathcal{M}_{ZZ}^L$ is repeated $r$ times to improve its reliability. In the first measurement we can time the preparation of the unprotected qubit and the cat state so that there are no storage errors prior to step in which the cat state interacts with the data. But in subsequent measurements storage errors on the data may accumulate during the $8r'$ time steps while the cat state is prepared for the next round. To estimate $\Perr^*(\Meas_X)$, we note that each of the $m+1$ cat-state qubits is acted upon by two CZ gates during each of $r'$ rounds of cat-state syndrome measurement, where a nondiagonal fault in any of these gates could result in a Pauli-frame $X$ error. Furthermore, a storage fault acting on the unprotected qubit, a fault in final $X$ measurement of the unprotected qubit, or a fault in the $CZ$ gate that couples the unprotected qubit to the cat state could cause $\mathcal{M}_{X}$ to fail. Therefore we find
\begin{widetext}
\begin{equation}
	\Perr^*(\Meas_X) \leq (r+1)(\eps + \eps')+ 8(r-1)r'(\eps_s + \eps'_s)+ 2 (m+1) rr' \eps'.
\end{equation}
\end{widetext}

\noindent A nondiagonal error acting on any data qubit can flip the outcome of $\Meas_{Z Z}^L$. Otherwise, a diagonal fault in each of at least $(r+1)/2$ measurement rounds could cause the $\Meas_{Z Z}^L$ to fail. Since in each round there are $m+1$ gates coupling the qubits to the data, $(m+1)r'$ CZ gates in the cat state preparation, as well as $m+1$ single-qubit preparations and measurements, we find
\begin{widetext}
\begin{equation}
	\Perr^*(\Meas_{Z Z}^L) \leq  r \eps' +8r'(r-1)\eps'_s +  m (2 r_+ + r) \eps' + \binom{r}{\frac{r+1}{2}} [(m + 1)(r' + 3 )(\eps + \eps')]^{(r+1)/2}.
\end{equation}
\end{widetext}

\begin{figure}[ht]
\begin{center}
\includegraphics[width=0.45\textwidth]{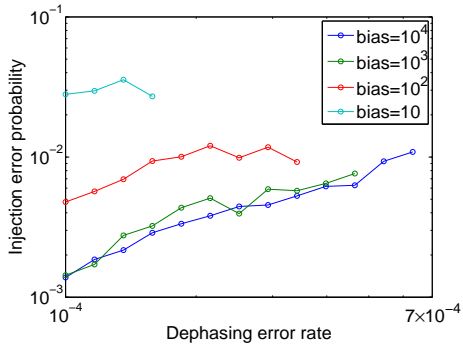}
\end{center}
\caption{\label{fig:inject-err}(Color online) Injection error probability for the optimal geometrically local Bacon-Shor codes found in \autoref{fig:encoded-err-local}, with the probability of an error on the unprotected state $\ket{\psi}$, $\eps_{\psi}$, set to be equal to the dephasing error probability $\eps$. }
\end{figure}

\autoref{fig:inject-err} bounds the injection error probability assuming the values of $m$, $r$ and $r'$ chosen to optimize the geometrically local CNOT gadget. For large biases the injection error probability is around 1\%, well below the thresholds for successful distillation, as will be seen in Sec.~\ref{sec:distillation}.

By choosing $r=1$ (no repetition of the $ZZ^L$ measurement), we can avoid storage locations, and  the leading contribution to $\Perr(\Inject)$ linear in $\eps$ (assuming $\eps' \ll \eps$) is  
\begin{equation}
\Perr(\Inject) = 3\eps + (m+1)(r'+3)\eps + \cdots.
\end{equation}
Increasing to $r=3$ entails increasing the sensitivity to storage errors, while providing better protection against diagonal errors in CZ gates:
\begin{equation}
\Perr(\Inject) = 5\eps + 16r'\eps_s + \cdots,
\end{equation}
which might be a significant improvement if $\eps_s \ll \eps$.

The probability of error in state injection depends on $m$ and also on $r'$, the number of times the cat-state syndrome measurement is repeated, which we assume matches the value of $r'$ in the CNOT$^L$ gate that follows the state injection step. We note, though, that the value of $r'$ might increase gradually as state injection proceeds. If there are multiple rounds of state distillation, we might be willing to accept a larger CNOT$^L$ error rate $\Perr(\mbox{CNOT$^L$})$ in early rounds where the error in the distilled state is higher, with $\Perr(\mbox{CNOT$^L$})$ declining in later rounds as the state's purity improves. Adjustment of the Bacon-Shor block size, and hence of $\Perr(\mbox{CNOT$^L$})$, is easy to incorporate in our circuit constructions, as there is no need for the control and target blocks in the CNOT$^L$ circuit to be of equal size. Using a smaller code in early rounds may save on overhead, but more importantly reducing the value of $m$ and $r'$ used in the first round improves the error $\Perr(\Inject)$ of the initially injected state. However, for simplicity, we did not invoke this strategy of gradually increasing the block size in the analysis of state distillation reported in Sec.~\ref{sec:distillation}.

\section{State distillation}\label{sec:distillation}

So far we have seen how, using asymmetric Bacon-Shor codes,  to perform fault-tolerant encoded versions of the operations in the ``CSS set'' $\mathcal{G}_{\rm CSS}^L$: the CNOT gate, preparations of the encoded states $|0\rangle^L$ and $|+\rangle^L$, and the measurements $\Meas_{X}^L$, $\Meas_{Z}^L$. To perform fault-tolerant universal quantum computation, we will need in addition to prepare high-fidelity encoded versions of the states
\begin{eqnarray}
|{+i}\rangle &=& \frac{1}{\sqrt{2}}\left( |0\rangle + i|1\rangle\right),\nonumber\\
|T\rangle &=& \frac{1}{\sqrt{2}}\left( |0\rangle + e^{i\pi/4}|1\rangle\right).
\end{eqnarray}
Using a $|+i\rangle$ ancilla state and CSS operations, we can teleport the Clifford group gates $Q=\exp\left(i\frac{\pi}{4} X\right)$ and $S=\exp\left(-i\frac{\pi}{4} Z\right)$, which suffice for generating the full Clifford group. Using a $|T\rangle$ ancilla, the $S$ gate, and CSS operations, we can teleport $T=\exp\left(-i\frac{\pi}{8} Z\right)$, completing a universal gate set. 

To prepare these encoded ancilla states, we first prepare noisy versions of the encoded states, and then use a distillation protocol to generate the needed high-fidelity versions of these states \cite{Bravyi08}. To distill the $|{+i}\rangle$ ancillas we need only CSS operations, and can use a circuit based on the [[7,1,3]] Steane code; it takes 7 noisy $\ket{{+}i}$ ($\ket{{-}i}$) ancillas and produces one clean $\ket{{-}i}$ ($\ket{{+}i}$) ancilla. The $|T\rangle$ distillation protocol uses both CSS operations and $|{+i}\rangle$ ancillas and is based on Reed-Muller codes \cite{Bravyi08}; it produces one clean $\ket{T}$ ancilla from 15 noisy ones. Other, more efficient, distillation protocols have been proposed recently \cite{Meier12,Bravyi12,Jones12}.

\begin{figure}[ht]
\begin{center}
\includegraphics[width=0.45\textwidth]{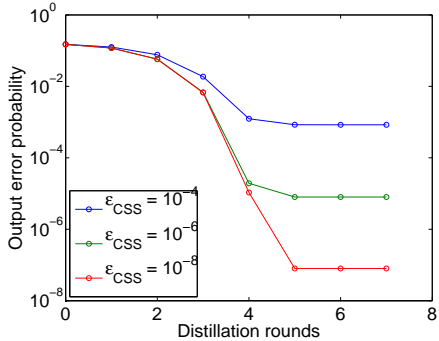}
\end{center}
\caption{\label{fig:T-distill-err}(Color online) Output error probability after $n$ levels of $\ket{T}$ distillation, starting from $\epsin = 0.15$, for various values of $\epsCSS$. The $\ket{{+}i}$ ancillas are assumed to be maximally distilled, so that $\eps_{\ket{{+}i}} = 4 \epsCSS $. Eventually the output probability levels off to around $8 \epsCSS$.}
\end{figure}

\begin{figure}[ht]
\begin{center}
\includegraphics[width=0.45\textwidth]{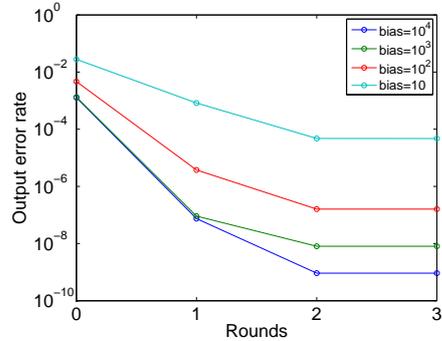}
\end{center}
\caption{\label{fig:T-inject-and-distill-err}(Color online) Output error probability after $n$ levels of $\ket{T}$ distillation, with the values of $\epsin$ and $\epsCSS$ taken from the optimal values of \autoref{fig:inject-err}, for $\eps = 10^{-4}$ and various biases. Only two rounds of distillation are required in each case to reach the floor.}
\end{figure}
In the original analysis \cite{Bravyi08} of the performance of these state distillation protocols, it was assumed that the CSS operations were perfect. Here we want to consider the case in which the CSS operations, protected by an optimally chosen Bacon-Shor code, are themselves noisy. We provide more details about our analysis in \cite{Brooks12}; state distillation using noisy CSS operations has also been discussed previously in, for example, Refs. \cite{raussendorf2007topological,fowler2009high,jochym2012robustness}.

When the CSS operations are perfect, we can attain ancilla states with arbitrarily low error by performing a sufficient number of rounds of distillation, assuming that the fidelity of the initial ancilla with the desired ideal state exceeds a threshold value. However, when the CSS operations are noisy, the error rate $\epsCSS$ for the CSS operations sets a nonzero floor on the error in the output ancilla states --- once this floor is reached, further rounds of distillation produce no further improvement, because the error in the output is dominated by the errors in the distillation circuit rather than the errors in the input ancilla states. We have designed fault-tolerant distillation circuits that minimize this noise floor, and have found the corresponding lower bounds on the output error for distillation of both $\ket{{+}i}$ and $\ket{T}$ states \cite{Brooks12}. For $\ket{{+}i}$ distillation, the best achievable error rate for our procedure is
\begin{equation}
	\eps_\ket{{+}i} \geq 4 \epsCSS + O(\epsCSS^2) ,
\end{equation} 
and for $\ket{T}$ distillation it is 
\begin{equation}
	\eps_\ket{T} \geq 8 \epsCSS + O(\epsCSS^2) .
\end{equation} 

A single round of the noisy distillation protocol improves the ancilla error probability substantially, provided that the CSS error rate $\epsCSS$ in the distillation circuit is small compared to the input error probability $\epsin$ of the noisy ancilla states, and also that $\epsin$ is safely below the (quite high) threshold error rate for the ideal distillation protocol. The convergence of the ancilla error rate to the floor $\approx 8 \epsCSS$ is illustrated for $\ket{T}$ distillation in \autoref{fig:T-distill-err}, for the case where the error rate of the input ancillas is $\epsin=0.15$. 

For initial ancillas with errors that are very close to the threshold of the ideal distillation protocol, it can in principle take many rounds of distillation to reach the floor set by the CSS error rate. On the other hand, if the initial ancillas are not too noisy, then it will take just a few rounds of distillation to reach this floor. In \autoref{fig:T-inject-and-distill-err}, the error rate of the output ancilla is plotted as a function of the number of rounds of distillation, where the input error rate is the injection error probability estimated in \autoref{fig:inject-err}. For the parameters shown, only two rounds of distillation suffice to reach the floor set by the logical CSS error rate. This relatively modest overhead cost for executing non-Clifford logical quantum gates is one of the advantages of our fault-tolerant scheme. 


\acknowledgments
This work was supported in part by the Intelligence Advanced Research Projects Activity (IARPA) via Department of Interior National Business Center contract number D11PC20165. The U.S. Government is authorized to reproduce and distribute reprints for Governmental purposes notwithstanding any copyright annotation thereon. The views and conclusions contained herein are those of the author and should not be interpreted as necessarily representing the official policies or endorsements, either expressed or implied, of IARPA, DoI/NBC
or the U.S. Government.
We also acknowledge support from NSF grant PHY-0803371, DOE grant DE-FG03-92-ER40701, and NSA/ARO grant W911NF-09-1-0442. The Institute for Quantum Information and Matter (IQIM) is an NSF Physics Frontiers Center with support from the Gordon and Betty Moore Foundation. 


\bibliographystyle{utphys}
\bibliography{bacon-shor-refs}


\end{document}